\documentclass[11pt]{article}
\pdfoutput=1

\usepackage{bm,amssymb,amsmath,mathrsfs}
\usepackage[usenames]{color}

\usepackage{dcolumn}
\usepackage[pdftex]{graphicx}

\newcommand{\bc}{\begin{center}}
\newcommand{\ec}{\end{center}}
\newcommand{\bit}{\begin{itemize}}
\newcommand{\eit}{\end{itemize}}
\newcommand{\bq}{\begin{equation}}
\newcommand{\eq}{\end{equation}}

\begin{document}

\baselineskip 3.7ex



\title{
\bf
Fits to data for a stored uncooled polarized deuteron beam
}

\vskip 0.2in
\author{
S.~R.~Mane \\ \; \\
Convergent Computing, Inc., P.O.~Box 561, Shoreham, NY 11786, USA
}

\vskip 0.2in
\date{\today}

\maketitle

\begin{abstract}
I perform tracking simulations to fit various measurements of the polarization for a stored uncooled 
polarized deuteron beam, published in the recent paper by Benati et.~al
(P.~Benati {\em et al.},  {\it Phys.~Rev.~ST Accel.~Beams} {\bf 15}, 124202 (2012)).
The collaboration kindly sent me datafiles of the polarization measurements,
and also pertinent details of the experimental data acquisition procedure.
The latter are essential to obtain quantitative fits to the data.
I describe my findings and inferences from the data.
In some cases I offer alternative interpretations of the data from that given by Benati et.~al.
I also correct some mistakes in my recent paper (S.~R.~Mane, {\em Nucl.~Inst.~Meth.} {\bf 726} 104--112 (2013)).
\end{abstract}

\vskip 0.25in
\noindent
{\em keywords}: 
storage rings, 
polarized beams, 
spin dynamics,
rf solenoid


\vfill\pagebreak
\section{\label{sec:intro} Introduction}

\begin{figure}[!htb]
\centering
\includegraphics[width=0.75\textwidth]{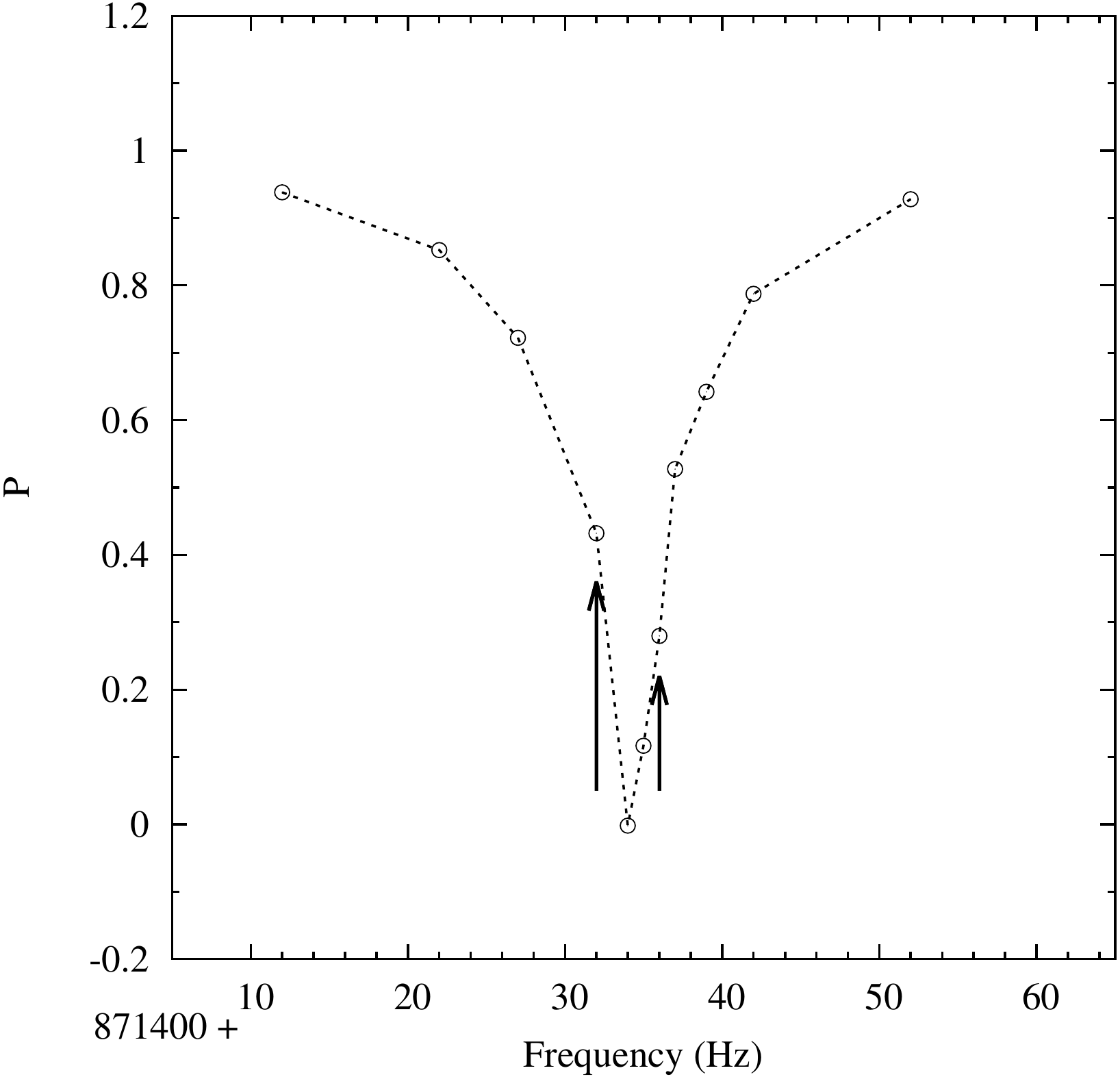}
\caption{\small
\label{fig:Fig22-data}
Data points published in Fig.~22 in \cite{Benati_etal_2012}.
The numerical values are given in Table \ref{tb:datafig22}.
The dotted line is just to guide the eye.
The arrows indicate the points at frequencies of 871432 Hz and 871436 Hz.
}
\end{figure}

In a recent paper, Benati et.~al.~\cite{Benati_etal_2012}
presented results for spin resonances for a stored polarized deuteron beam, induced by an rf solenoid.
The effects of synchrotron oscillations on the spin precessions were found to be significant.
I published a recent paper \cite{ManeSyncTunMod2013} 
deriving analytical formulas for the synchrotron tune modulation of spin resonances
induced by a localized rf solenoid or rf dipole.
I published various analyzes of the data in \cite{Benati_etal_2012} in my paper \cite{ManeSyncTunMod2013} .
In this paper, I shall present more detailed
theoretical simulations to analyze the polarization measurements for the uncooled beam 
in the recent paper by Benati et.~al.~\cite{Benati_etal_2012}.
(I explained in \cite{ManeSyncTunMod2013} 
that the effects of the synchrotron tune modulation 
on the spin precessions for the cooled beam in \cite{Benati_etal_2012} were negligible,
and the data in \cite{Benati_etal_2012} for the cooled beam could be fitted using a monochromatic beam.)
The collaboration kindly sent me datafiles of the polarization measurements 
for the various data points in the resonance dip for the uncooled beam in 
Fig.~22 in Benati et al.~\cite{Benati_etal_2012}.
The (frequency, polarization) values are displayed in Table \ref{tb:datafig22}.
(Note that `polarization' will always mean `normalized polarization' below.)
A graph of the data is plotted in Fig.~\ref{fig:Fig22-data}.

I was able to fit the data using my own tracking simulations. 
However, to do so I had to understand the experimental procedure of the measurements,
because there were some significant details I had not understood from a reading of \cite{Benati_etal_2012}.
I thank the collaboration for explaining the experimental procedure to me.
The following details are significant:

\bit
\item
In response to a query 
about the contents of some of the datafiles,
the collaboration kindly sent me updated files with improved normalization;
it is these values which are tabulated in Table \ref{tb:datafig22}.
The difference with the values plotted in 
Fig.~22 in Benati et al.~\cite{Benati_etal_2012}
is too small to discern visually.
I thank the collaboration for responding courteously to my query.
{\em N.B.:} The revised normalizations apply only to the data in Figs.~21 and 22 of \cite{Benati_etal_2012}.
The term `data' will always mean `revised data' for the above cases.
I shall also present fits to the data in Figs.~12, 16 and 17 of \cite{Benati_etal_2012};
in those cases the data are the same as in \cite{Benati_etal_2012}.

\item
It is stated in \cite{Benati_etal_2012} that the rf solenoid was ramped linearly to full strength in 200 ms,
and that the measured polarization depended on the ramp rate.
I confirmed this in my simulations. 

\item
I found that a single value for the resonance center would not fit all the points.
The center of the resonance is given in \cite{Benati_etal_2012} at 871434 Hz.
However, the two data points at 871432 Hz and 871436 Hz,
indicated by the arrows in Fig.~\ref{fig:Fig22-data},
which should be equidistant from the center of the resonance,
do {\em not} have equal polarizations.
I found that some points were fitted by setting the resonant frequency to 871434.0 Hz
and the rest using 871434.4 Hz.
(It is commented in \cite{Benati_etal_2012} that the points do not all seem
to correspond to the same resonance location; see below.)

\item
I ran tracking simulations using resonance centers of $f_{\rm res} = 871434.0$ Hz and $871434.4$ Hz.
Even so, I was unable to fit all the points, viz.~the three leftmost and the rightmost point in 
Fig.~22 in \cite{Benati_etal_2012}.
See Fig.~\ref{fig:sweep-uncooled-1}, to be explained below.
My simulation results were sufficiently precise that
I realized something had to be different about the experimental parameter settings 
when measuring these four points; they do {\em not} belong on the same resonance curve as the rest.
However, at this stage it is impossible to offer a definitive reason why.
I offer my hypothesis below; see Fig.~\ref{fig:sweep-uncooled-2}, which I shall explain below.

\item
I shall also fit the data in Figs.~12, 16, 17 and 21 in \cite{Benati_etal_2012},
and will discuss them below.

\eit

The structure of this paper is as follows.
Sec.~\ref{sec:fit} describes general properties of my fitting procedure.
Sec.~\ref{sec:corr} describes some corrections to my recent paper \cite{ManeSyncTunMod2013}.
Sec.~\ref{sec:retro} presents some remarks on the use of various distributions of the particle orbits to fit the data.
Sec.~\ref{sec:sim} presents my tracking simulations for the resonance dip of the uncooled beam in Fig.~22 in \cite{Benati_etal_2012}.
Sec.~\ref{sec:syncosc} presents a comparison with some analytical formulas I derived in \cite{ManeSyncTunMod2013}.
Sec.~\ref{sec:fig21171216} presents my tracking simulations for the data in
Figs.~12, 16, 17 and 21 in \cite{Benati_etal_2012}.
Sec.~\ref{sec:conc} concludes.

\vfill\pagebreak
\section{\label{sec:fit} Fits to data}
I shall present my detailed investigations later.
To summarize: 
\bit
\item
Of the eleven points displayed in Fig.~22 in \cite{Benati_etal_2012},
five were fitted with resonant frequency of $f_{\rm res} = 871434.4$ Hz and
and six were fitted with resonant frequency of $871434.0$ Hz.

\item
I employed an rf solenoid `resonance strength' of $\varepsilon_{\rm FWHM} = 2.66 \times 10^{-5}$,
as given in \cite{Benati_etal_2012}.
The rf solenoid field amplitude was ramped linearly to full strength in a time $t_{\rm ramp} = 0.2$ s.
It was essential to include this ramp in my simulations, to obtain a quantitative fit to the data.

\item
All of my tracking simulations were computed using a Gaussian distribution of the particle orbits.
I employed an r.m.s.~relative momentum spread of $\sigma_p = 8.02\times 10^{-4}$,
which is the value stated in \cite{Benati_etal_2012} for the uncooled beam.
The initial value of the synchrotron oscillation phase was distributed uniformly in $[0,2\pi)$.

\item
However, there were four ponts which I could not fit with the above simulation parameters.
These were the three leftmost points and the rightmost point in Fig.~22 in \cite{Benati_etal_2012}.
The results of my tracking simulations were sufficiently precise that I ruled out statistical fluctuations,
even though my numerical work consisted of Monte Carlo simulations, and the data were themselves statistical samples.
Something must have been different about the experimental settings when measuring these four points.
I was able to fit these four points using a smaller r.m.s.~relative momentum spread of $\sigma_p = 6.0\times 10^{-4}$.
This is simply a hypothesis; it is a possible but not conclusive explanation of the data;
there were most likely multiple causes.

\item
I therefore compiled the data into a `common set' where all the points were based on a resonance center of 871434.0 Hz.
I did this by shifting the frequency down by $0.4$ Hz, for all the rows in Table \ref{tb:datafig22comment}
for which $f_{\rm res} = 871434.4$ Hz, so that the effective resonance center was 871434.0 Hz for all the points.
The resultant dataset is tabulated in Table \ref{tb:datafig22modified},
where column 1 is labeled `effective' frequency.
Column 2 displays a key to indicate a frequency shift (`*') 
or a reduced r.m.s.~relative momentum spread (`--').
I then plotted the modified values in Fig.~\ref{fig:sweep-uncooled-2}:
\bit
\item
Squares: points for which $f_{\rm res} = 871434.4$ Hz.
\item
Circles: points for which $f_{\rm res} = 871434.0$ Hz.
\item
Triangles: fitted using $\sigma_p = 6.0\times 10^{-4}$
($f_{\rm res} = 871434.0$ Hz for all of them).
\eit
I ran two tracking simulations, both computed using a resonant frequency of 871434.0 Hz,
using $\sigma_p = 8.02\times 10^{-4}$ and $6.0\times 10^{-4}$.
The respective outputs are displayed as the solid and dashed curves in Fig.~\ref{fig:sweep-uncooled-2}.
The fit is almost perfect:
the four points with $\sigma_p = 6.0\times 10^{-4}$ (triangles) all lie on the dashed curve, and
the seven points with $\sigma_p = 8.02\times 10^{-4}$ (five squares and two circles) all lie on the solid curve.

\item
{\em Alternative hypothesis:}\\
I actually employed a different hypothesis in my first attempt to fit the 
four `triangle points' in Table \ref{tb:datafig22modified}.
I initially assumed $\sigma_p = 8.02\times 10^{-4}$ for all the points.
I found that the four `triangle points' could all be fitted using a reduced ramp time of 0.02 s.
The outputs are displayed as the solid and dotted curves in Fig.~\ref{fig:sweep-uncooled_ramp}.
The data points are the same as in Fig.~\ref{fig:sweep-uncooled-2}, also the solid curve,
but the dotted curve was computed using $t_{\rm ramp} = 0.02$ s
(and the full r.m.s.~relative momentum spread $\sigma_p = 8.02\times 10^{-4}$).
However, I was informed that the ramp time was 0.2 s for all of the points displayed in 
Fig.~22 in Benati et al.~\cite{Benati_etal_2012},
so this hypothesis was not a valid explanation.

\item
The fits to other figures in \cite{Benati_etal_2012} all employed a Gaussian distribution with
$\sigma_p = 8.02\times 10^{-4}$, and will be described below.

\eit

\vfill\pagebreak
\section{\label{sec:corr} Corrections to \cite{ManeSyncTunMod2013}}

I published a recent paper \cite{ManeSyncTunMod2013},
in which I made various comments about the data and simulations by the authors in \cite{Benati_etal_2012}.
Some of my claims were incorrect and must be revised:

\bit
\item
In \cite{ManeSyncTunMod2013}, I fitted the points in Fig.~22 in \cite{Benati_etal_2012}
using a Lorentzian with $\varepsilon_{\rm FWHM} \simeq 1.0 \times 10^{-5}$.
However, that fit treated all the points as if they were part of a single resonance curve.
From information kindly supplied to me by the collaboration,
I now know that the resonant frequency was {\em not} the same for all the points.
Hence the data points in Fig.~22 in \cite{Benati_etal_2012}
do {\em not} all lie on a single resonance curve.
Hence it is inappropriate to fit the data using a single Lorentzian curve.
See my revised fits in
Figs.~\ref{fig:sweep-uncooled-1} and \ref{fig:sweep-uncooled-2}, as I explained above.

\item
In \cite{ManeSyncTunMod2013}, I claimed to fit the data points in Fig.~21 in \cite{Benati_etal_2012}
and to deduce that the data was 4 Hz off resonance.
The fit in \cite{ManeSyncTunMod2013} was made without a full understanding of the 
experimental data acquisition procedure.
The correct offset from the resonance center is 2.4 Hz, to be explained below.

\item
I also present improved fits for the data in Fig.~17 in \cite{Benati_etal_2012},
based on a better understanding of the data acquisition procedure.
My hypothesis for Fig.~17 in \cite{Benati_etal_2012}
is still the same, viz.~the beam was 0.025 Hz off resonance,
but I offer improved fits to justify my claim.

\eit

\vfill\pagebreak
\section{\label{sec:retro} Retrofitting}

I could of course fit every point in Fig.~22 in \cite{Benati_etal_2012}
exactly by retrofitting the r.m.s.~relative momentum spread for each point.
After all, the momentum spread would not have been exactly the same for every bunch.
However, such a procedure has no predictive value.
It is known that the use of a Gaussian distribution is an approximation, 
and cannot be taken seriously to the extent of retrofitting to individual points.

However, the authors adjust the distribution of the synchrotron oscillation amplitudes in \cite{Benati_etal_2012}:
\bit
\item
``The calculation shown in Fig.~21 [{\em of \cite{Benati_etal_2012}}]
represents a readjustment of the number of tracks for a selected set of synchrotron amplitudes.''

\item
``Besides sensitivity to the solenoid strength and the
ramping time, there is also a dependence on the distribution
of synchrotron amplitudes. Variations in the trend of
the reproduction may reflect changes in the amplitude
distribution for the uncooled case from run to run.''

\item
``The dependence of the average polarization on the synchrotron
amplitude distribution, the rf-solenoid strength,
and the ramping time makes these comparisons a strong
test of the simple ``no lattice'' model used here.''

\eit
I see no evidence to justify such a claim.
I could fit every point exactly using a Gaussian distribution and retrofitting the r.m.s.~relative momentum spread for each point.
I see no evidence in the data to support any specific model of the particle oscillation amplitudes.
Of course a Gaussian distribution of orbits is by itself a model, but this is typically the default model,
even though it is clearly an approximation.
However, I see no evidence in the data to indicate a systematic deviation from a Gaussian distribution of orbits.

I initially attempted to fit the four `triangle points' in Table \ref{tb:datafig22modified}
by reducing the ramp time to 0.02 s.
I was able to fit the points, but I discovered the experiment used a ramp time of 0.2 s for all the points.
I then fixed the ramp time at 0.2 s and searched for an alternative explanation for the
four `triangle points' in Table \ref{tb:datafig22modified}.
I found that reducing the r.m.s.~relative momentum spread could do the job,
as displayed in Fig.~\ref{fig:sweep-uncooled-2}.
This demonstrates that there are multiple ways to modify the simulation parameters to fit the 
four `triangle points' in Table \ref{tb:datafig22modified}.
It is impossible at this stage to determine a definitive explanation for these four points.
What I {\em can} say, with confidence, is that these four points do not belong on the same resonance curve as the rest.

\vfill\pagebreak
\section{\label{sec:sim} Tracking simulations}
\subsection{General}
The collaboration kindly sent me the data files of the polarization measurements for the individual points
for Fig.~22 in \cite{Benati_etal_2012}.
They later sent me revised files, and I shall work with the revised files below.
I shall display the results of tracking simulations to fit the individual points,
and the lessons I learned thereby.

\begin{figure}[!htb]
\centering
\includegraphics[width=0.75\textwidth]{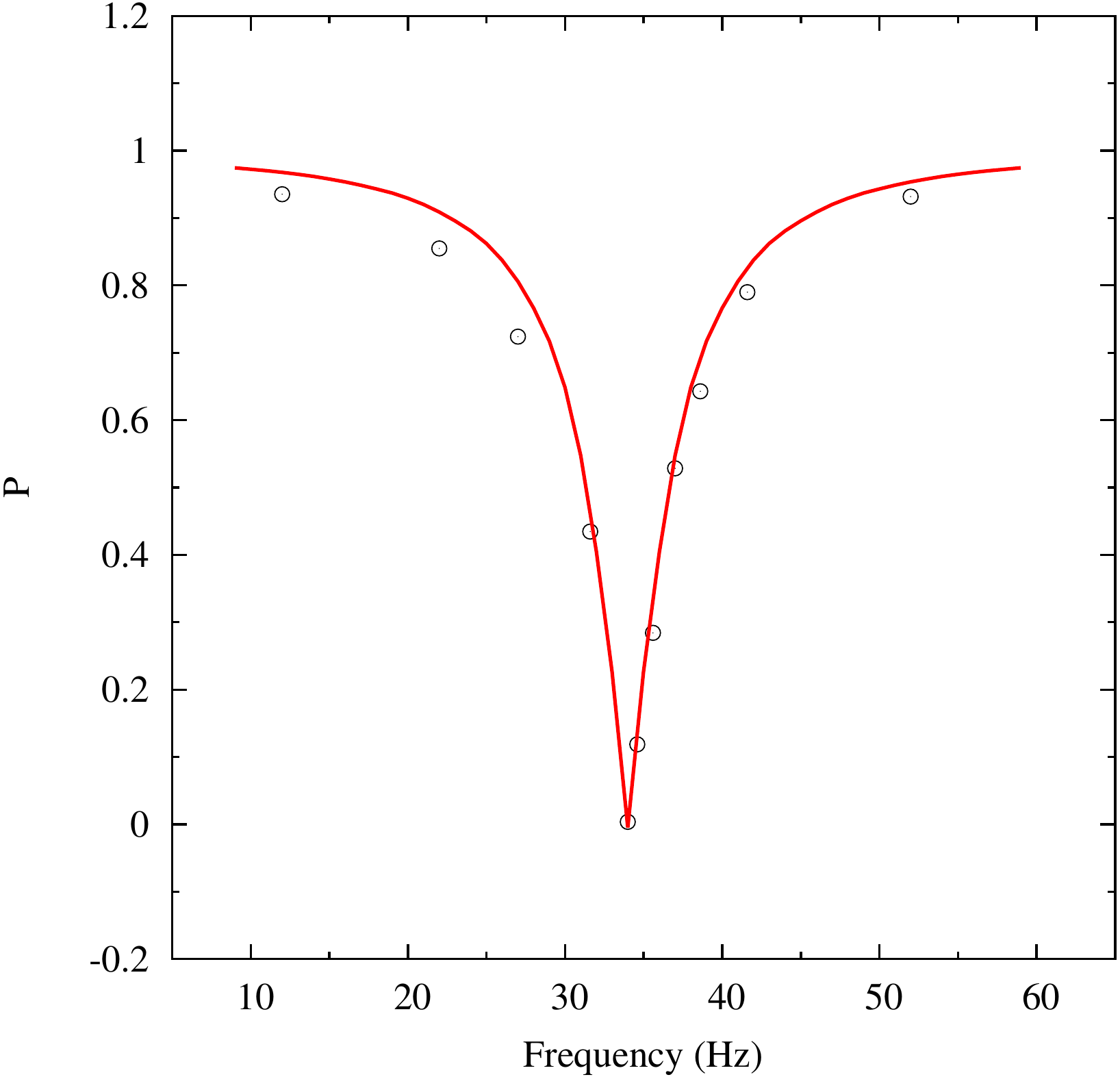}
\caption{\small
\label{fig:sweep-uncooled-1}
Plot of data points tabulated in Table \ref{tb:datafig22modified}, using `effective frequencies'
for some data points, as explained in the text.
The solid curve is the output of a tracking simulations using a resonant frequency of 871434.0 Hz.
The three leftmost points, and also the rightmost point, are not fitted by the curve;
this will be explained in the text.
}
\end{figure}

\begin{figure}[!htb]
\centering
\includegraphics[width=0.75\textwidth]{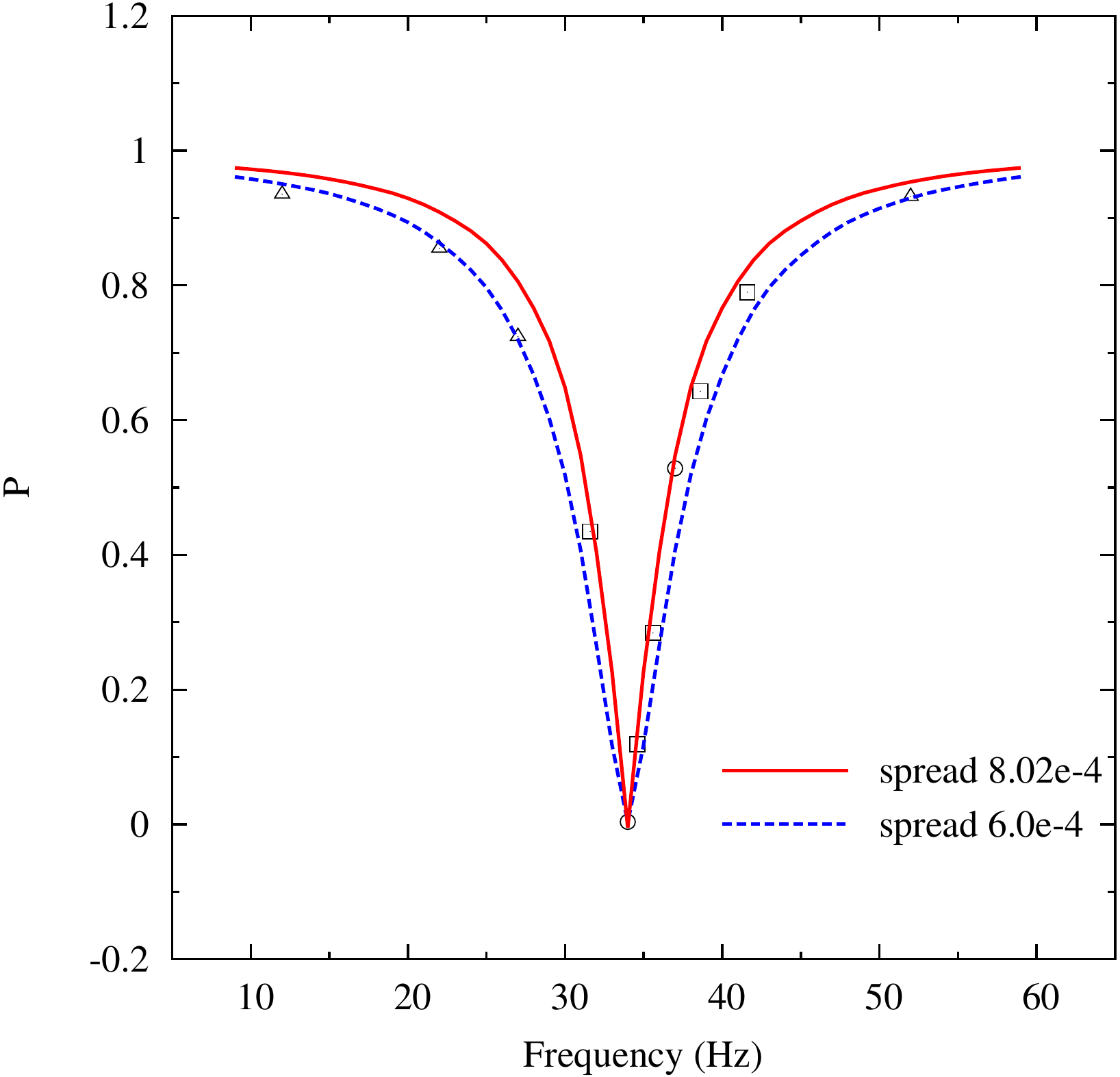}
\caption{\small
\label{fig:sweep-uncooled-2}
Plot of data points tabulated in Table \ref{tb:datafig22modified}, using `effective frequencies'
for some data points, as explained in the text.
The key for the data points is explained in Table \ref{tb:datafig22modified}.
The solid and dashed curves are the outputs of tracking simulations using 
r.m.s.~relative momentum spreads of $\sigma_p = 8.02 \times 10^{-4}$ (solid curve)
and $\sigma_p = 6.0 \times 10^{-4}$ (dashed curve).
}
\end{figure}

\begin{figure}[!htb]
\centering
\includegraphics[width=0.75\textwidth]{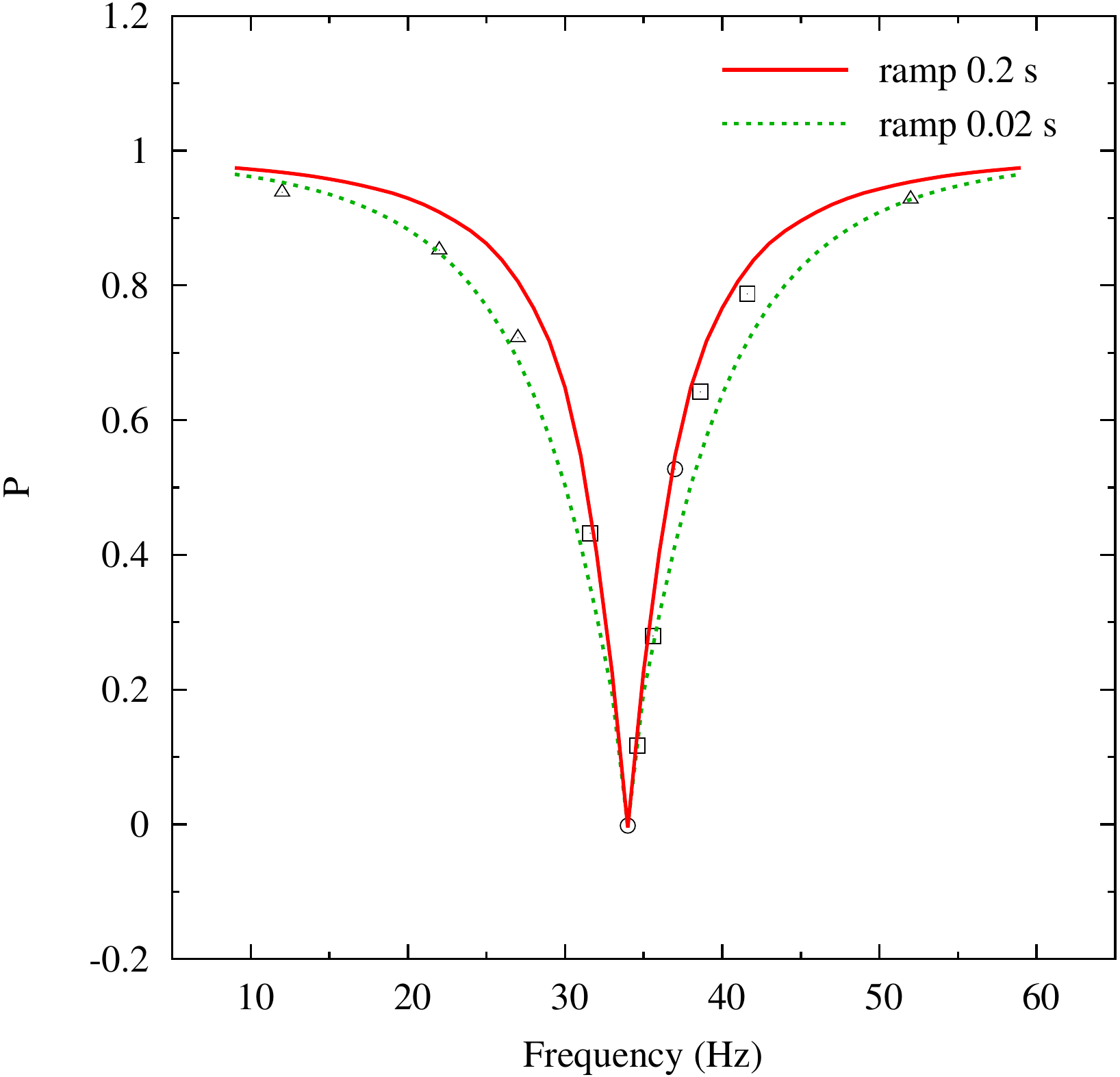}
\caption{\small
\label{fig:sweep-uncooled_ramp}
Plot of data points tabulated in Table \ref{tb:datafig22modified}, using `effective frequencies'
for some data points, as explained in the text.
The key for the data points is explained in Table \ref{tb:datafig22modified}.
The solid and dotted curves are the outputs of tracking simulations using ramp times of 0.2 s and 0.02 s, respectively.
(However, note that the actual ramp time in the experimental studies was 0.2 s for all the points.)
I employed an r.m.s.~relative momentum spread of $\sigma_p = 8.02 \times 10^{-4}$ in all cases.
}
\end{figure}

I employed the parameter values stated in \cite{Benati_etal_2012}.
I employed an rf solenoid `resonance strength' of $\varepsilon_{\rm FWHM} = 2.66 \times 10^{-5}$.
The rf solenoid field amplitude was ramped linearly to full strength in a time of 0.2 s.
I employed a Gaussian distribution of particle orbits with an r.m.s.~relative momentum spread of $\sigma_p = 8.02\times 10^{-4}$
and a uniform distribution in $[0,2\pi)$ for the intial phase of the synchrotron oscilllations.
The spins were all vertical at the start of a tracking simulation
and I measured the vertical polarization turn by turn.
When I plotted Figs.~\ref{fig:sweep-uncooled-1} and \ref{fig:sweep-uncooled-2},
I employed the {\em same} distribution of particle orbits to generate all the points in the solid curve.
(It is the same curve in Figs.~\ref{fig:sweep-uncooled-1}, \ref{fig:sweep-uncooled-2} and \ref{fig:sweep-uncooled_ramp}.)
I did the same for the dashed curve in Fig.~\ref{fig:sweep-uncooled-2}.
I employed same procedure for Fig.~\ref{fig:sweep-uncooled_ramp},
where I used the same set of orbits for both the solid and dotted curves.

The experimental procedure was as follows.
The rf solenoid was switched on at time $t=0$ (say) and ramped linearly to full strength, with $t_{\rm ramp} = 0.2$ s.
The rf solenoid was then operated at full strength for several seconds.
The vertical polarization was measured starting from $t=t_0=0.5$ s, for several seconds, say up to $t=T$.
The reported polarization in \cite{Benati_etal_2012} 
was the average vertical polarization from $t=t_0$ to $t=T$, i.e.
\bq
P_{\rm avg} = \frac{1}{T-t_0}\,\int^T_{t_0} P_{\rm vert}(t)\, dt \,.
\eq
The rf solenoid was ramped down to zero eventually, but this does not matter; 
the rf solenoid was at full strength in the averaging period $t_0 \le t \le T$.
I followed the above procedure in my tracking simulations.
I found that it was essential to include the initial ramp, to obtain a quantitative fit to the data.

\vfill\pagebreak
\subsection{Runs 86, 88, 89, 91 and 93: resonant frequency}
These are the points closest to the center of the resonance, hence I began with them.
The center of the resonance is given in \cite{Benati_etal_2012} at 871434 Hz.
However, the two data points in Runs 86 and 89, at 871432 Hz and 871436 Hz respectively,
should then be equidistant from the center of the resonance,
but I noticed that they do not have equal polarizations.
The polarization at 871432 Hz is approximately 0.43 and at 871436 Hz it is about 0.28.
My tracking simulations, using a resonant frequency of $f_{\rm res} = 871434.0$ Hz, 
yielded polarizations of about 0.36 at both points.
This indicated to me that the resonance center was in fact {\em not} at 871434.0 Hz;
it must be higher, so that the point at 871436 Hz is closer to the resonance center and hence has a lower polarization.
I guessed a shift of 0.5 Hz, i.e.~$f_{\rm res} = 871434.5$ Hz.
This worked well.
I obtained a better fit using an offset of 0.4 Hz, i.e.~$f_{\rm res} = 871434.4$ Hz,
which fitted all three data points in Runs 86, 88 amd 89, at 871432, 871435 and 871436 Hz, respectively.
However, the polarization was $0.0036147$ at 871434.0 Hz (Run 91, see Table \ref{tb:datafig22}),
hence the resonant frequency was obviously 871434.0 for this point.
I also found that $f_{\rm res} = 871434.0$ Hz gave a good fit for Run 93, at 871437 Hz.

\begin{figure}[!htb]
\centering
\includegraphics[width=0.75\textwidth]{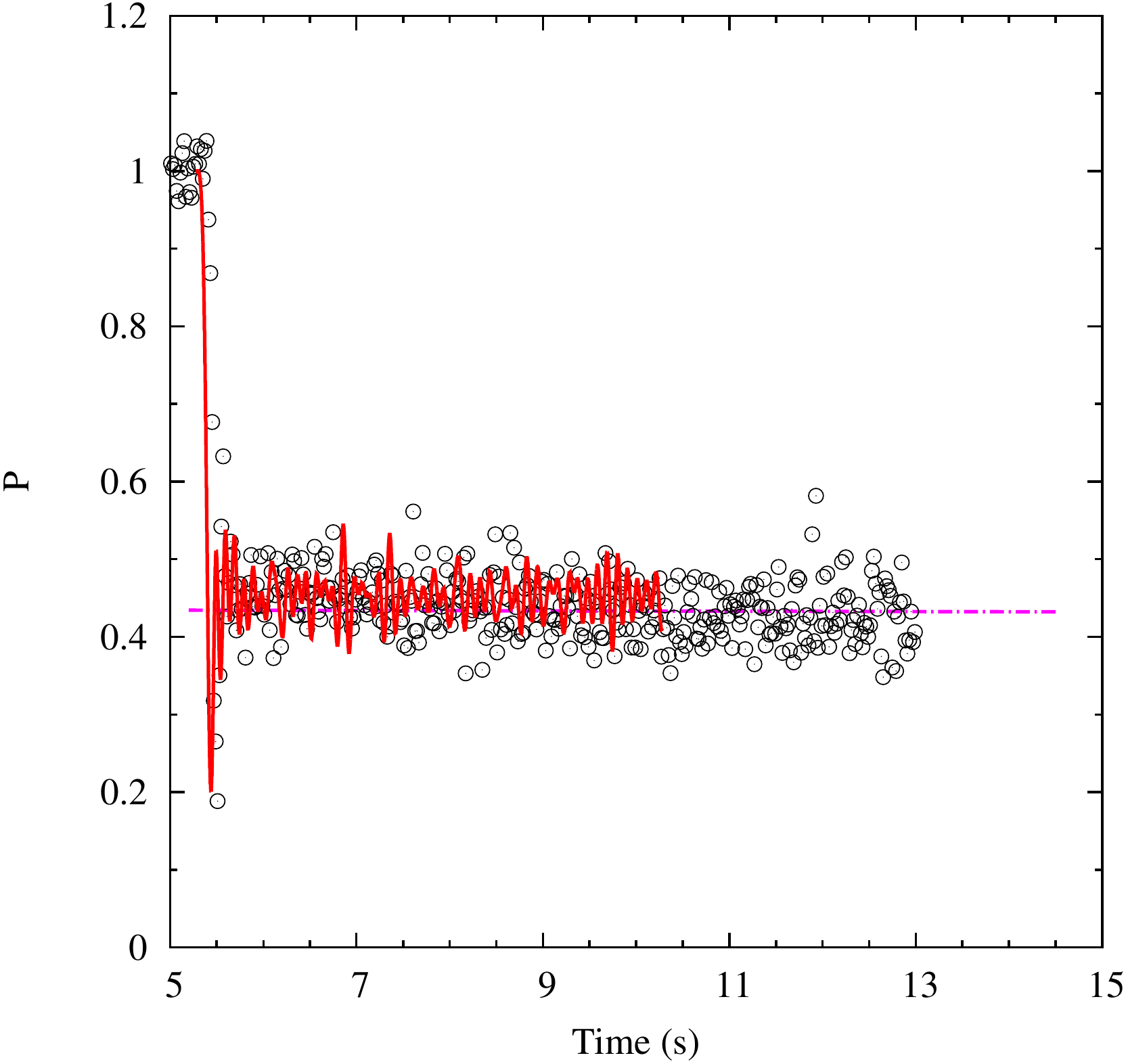}
\caption{\small
\label{fig:run86}
Plot of the data (circles) in the revised datafile sent to me by the collaboration for Run 86,
which is the data point at 871432 Hz in Fig.~22 in \cite{Benati_etal_2012}.
The solid curve is the output of a tracking simulation using a resonant frequency of 871434.4 Hz.
The dotdash line indicates a polarization of 0.43449, which is the revised value sent to me by the collaboration
(see Table \ref{tb:datafig22}).
}
\end{figure}

\begin{figure}[!htb]
\centering
\includegraphics[width=0.75\textwidth]{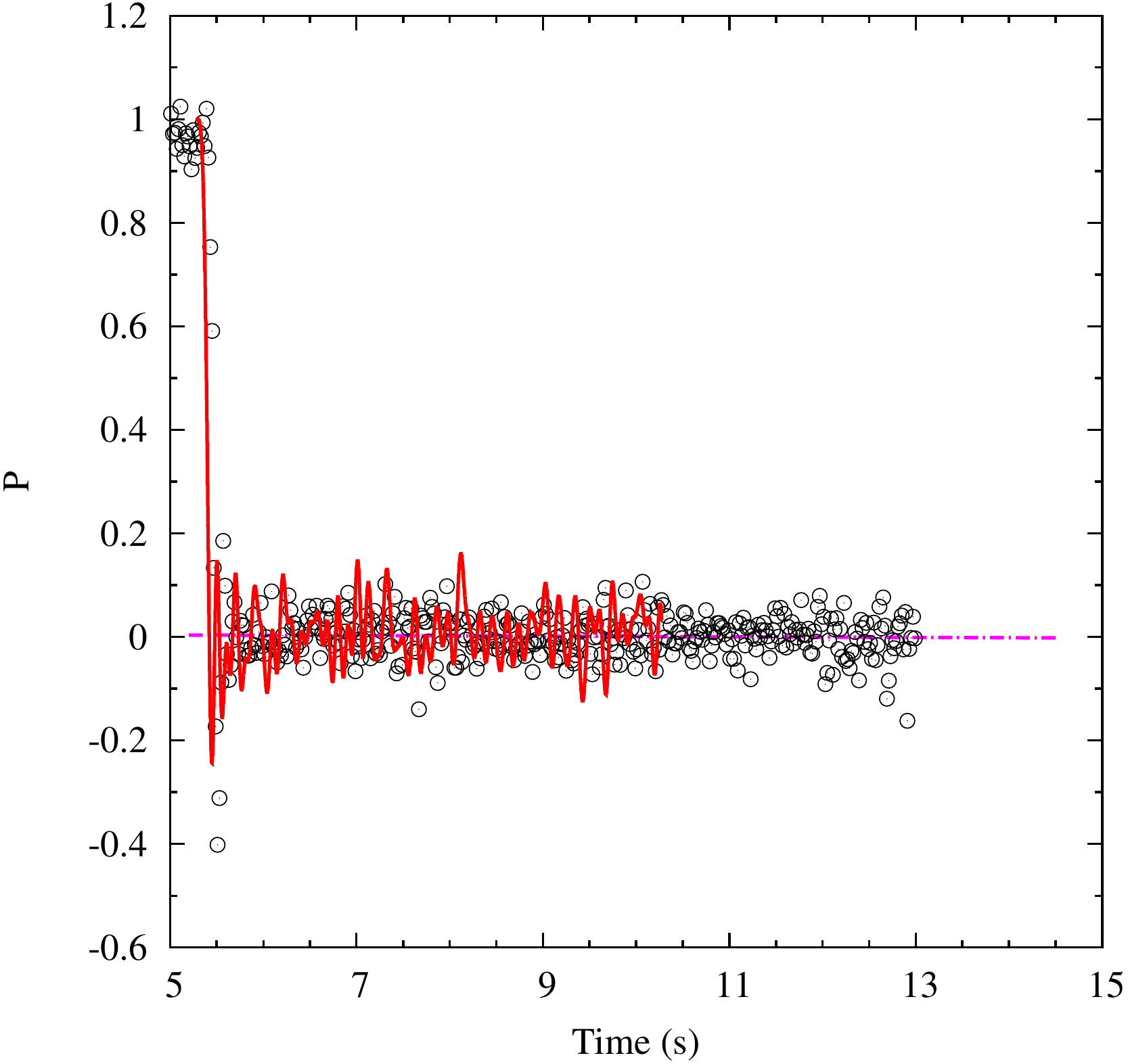}
\caption{\small
\label{fig:run91}
Plot of the data (circles) in the revised datafile sent to me by the collaboration for Run 91,
which is the data point at 871434 Hz in Fig.~22 in \cite{Benati_etal_2012}.
The solid curve is the output of a tracking simulation using a resonant frequency of 871434.0 Hz.
The dotdash line indicates a polarization of 0.0036147, which is the revised value sent to me by the collaboration
(see Table \ref{tb:datafig22}).
}
\end{figure}

\begin{figure}[!htb]
\centering
\includegraphics[width=0.75\textwidth]{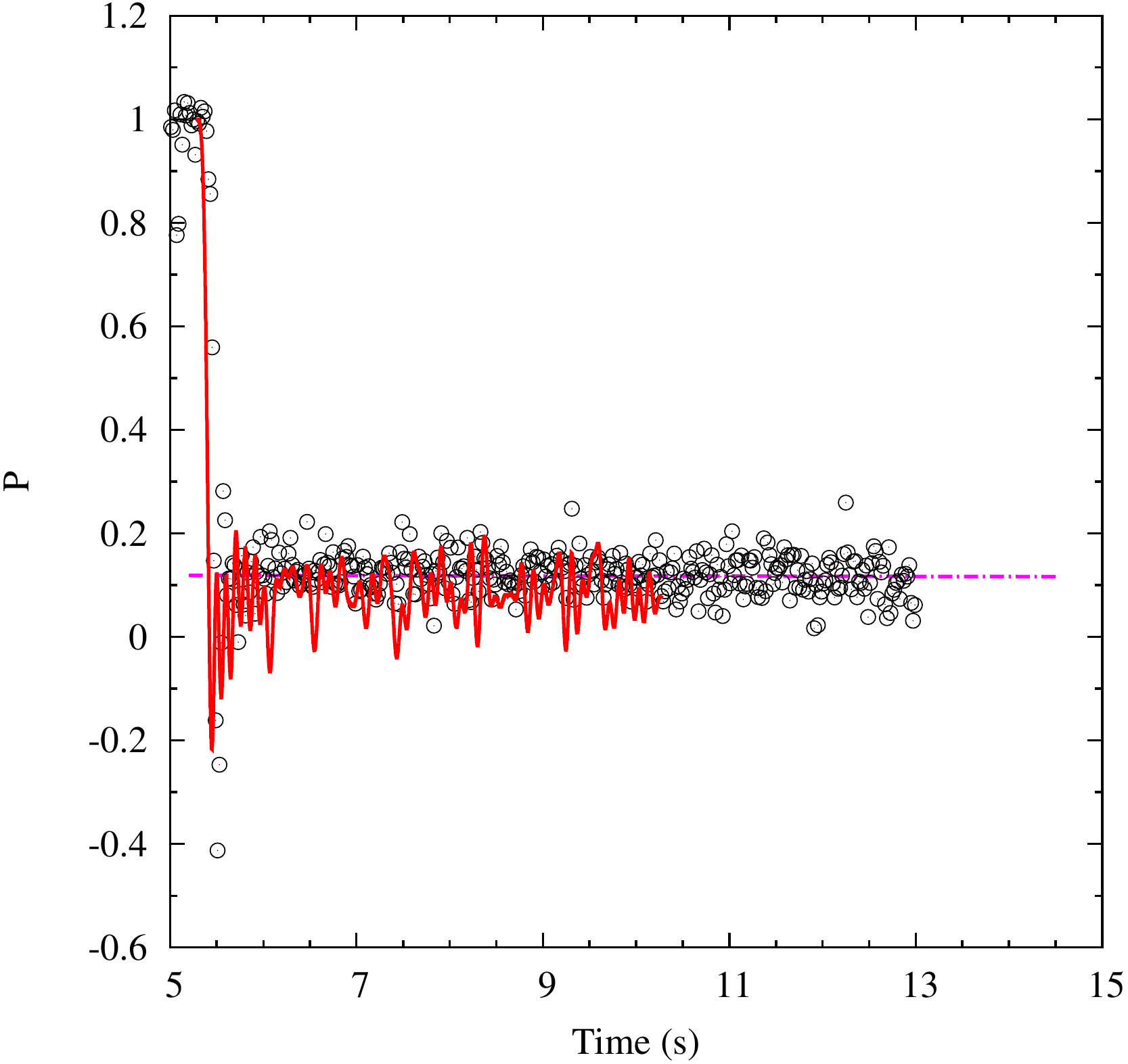}
\caption{\small
\label{fig:run88}
Plot of the data (circles) in the revised datafile sent to me by the collaboration for Run 88,
which is the data point at 871435 Hz in Fig.~22 in \cite{Benati_etal_2012}.
The solid curve is the output of a tracking simulation using a resonant frequency of 871434.4 Hz.
The dotdash line indicates a polarization of 0.11855, which is the revised value sent to me by the collaboration
(see Table \ref{tb:datafig22}).
}
\end{figure}

\begin{figure}[!htb]
\centering
\includegraphics[width=0.75\textwidth]{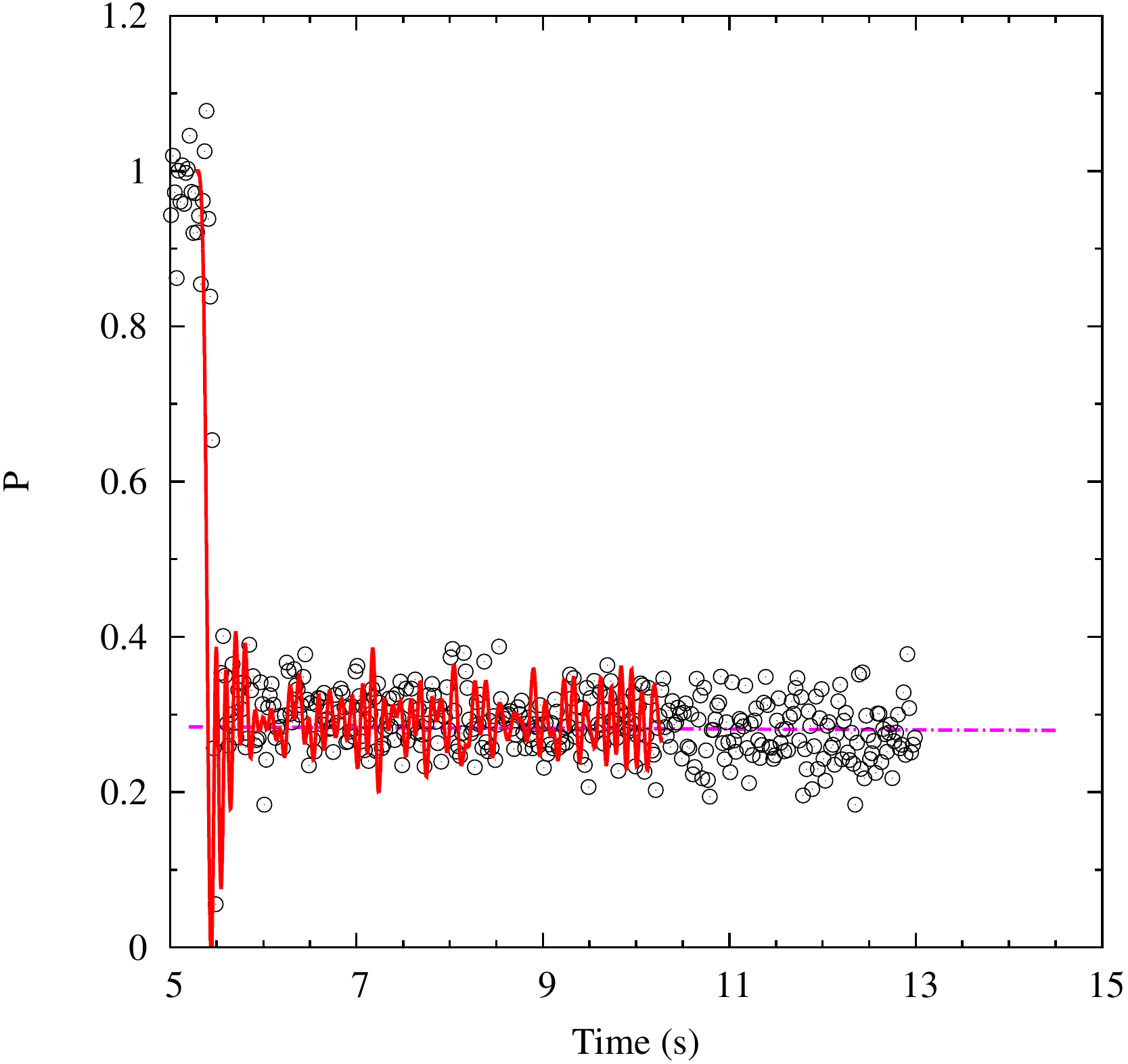}
\caption{\small
\label{fig:run89}
Plot of the data (circles) in the revised datafile sent to me by the collaboration for Run 89,
which is the data point at 871436 Hz in Fig.~22 in \cite{Benati_etal_2012}.
The solid curve is the output of a tracking simulation using a resonant frequency of 871434.4 Hz.
The dotdash line indicates a polarization of 0.28409, which is the revised value sent to me by the collaboration
(see Table \ref{tb:datafig22}).
}
\end{figure}

\begin{figure}[!htb]
\centering
\includegraphics[width=0.75\textwidth]{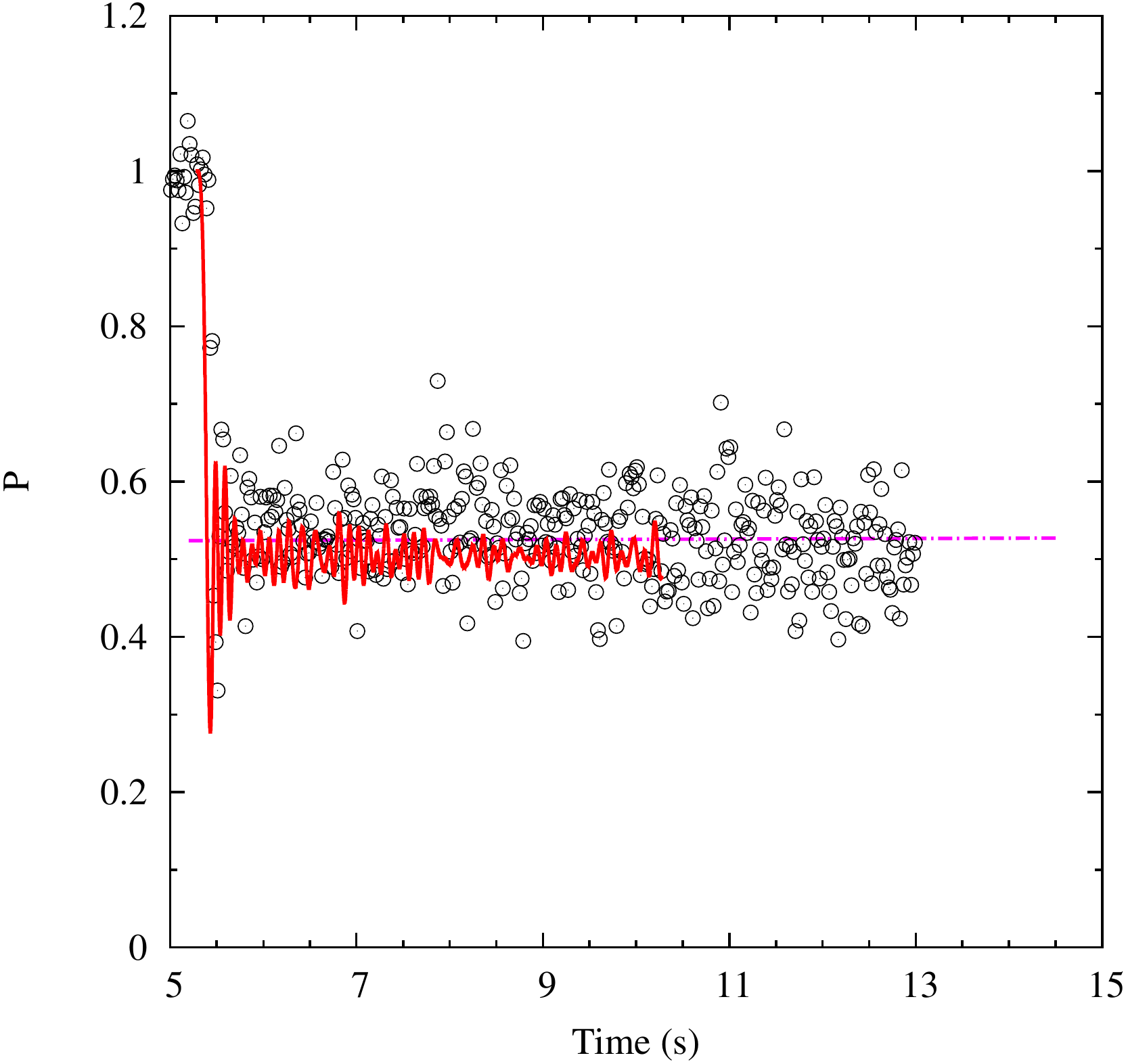}
\caption{\small
\label{fig:run93}
Plot of the data (circles) in the revised datafile sent to me by the collaboration for Run 93,
which is the data point at 871437 Hz in Fig.~22 in \cite{Benati_etal_2012}.
The solid curve is the output of a tracking simulation using a resonant frequency of 871434.0 Hz.
The dotdash line indicates a polarization of 0.52834, which is the revised value sent to me by the collaboration
(see Table \ref{tb:datafig22}).
}
\end{figure}

I display plots of the data and my simulation results in Figs.~\ref{fig:run86}--\ref{fig:run93}.
The plots are arranged in ascending order of the rf solenoid frequency, i.e.~Runs 86, 91, 88, 89 and 93.
In all cases, the data are plotted as circles and the solid curve plots the output from my tracking simulation.
The dotdash line is the revised average polarization level sent to me by the collaboration
(see Table \ref{tb:datafig22}).
Overall, the simulation results fit the data well.

\vfill\pagebreak
\subsection{Runs 85 and 90}
I subsequently fitted the data for runs 85 and 90, at 817442 Hz and 817439 Hz, respectively, using $f_{\rm res} = 871434.4$ Hz. 
These points are farther from the resonance center, and did not serve to establish the resonant frequency.
I had learned by now that most of the data points could be fitted using one of two resonant frequencies,
viz.~817434.0 Hz and 817434.4 Hz.

\begin{figure}[!htb]
\centering
\includegraphics[width=0.75\textwidth]{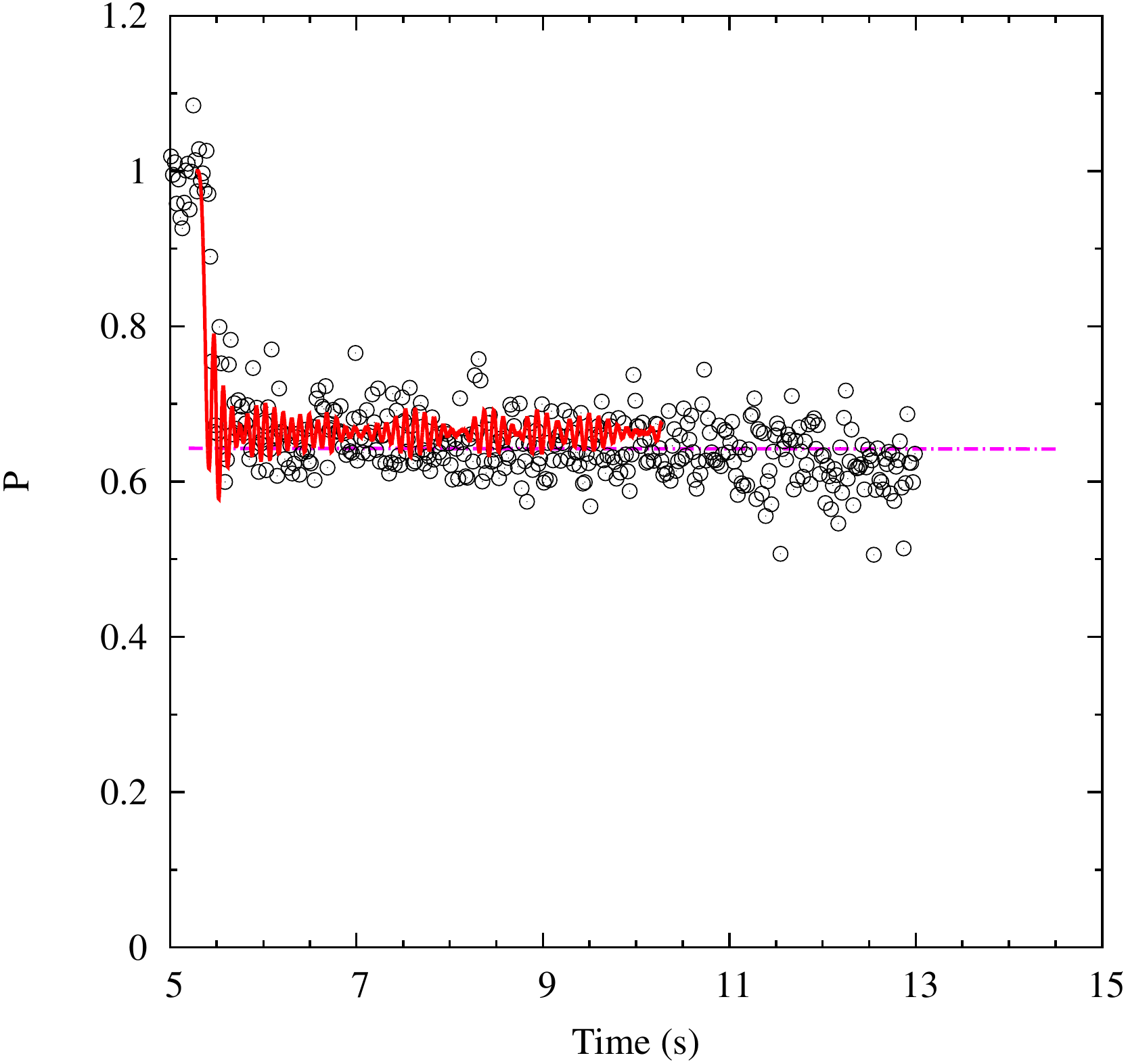}
\caption{\small
\label{fig:run90}
Plot of the data (circles) in the revised datafile sent to me by the collaboration for Run 90,
which is the data point at 871439 Hz in Fig.~22 in \cite{Benati_etal_2012}.
The solid curve is the output of a tracking simulation using a resonant frequency of 871434.4 Hz.
The dotdash line indicates a polarization of 0.64286, which is the revised value sent to me by the collaboration
(see Table \ref{tb:datafig22}).
}
\end{figure}

\begin{figure}[!htb]
\centering
\includegraphics[width=0.75\textwidth]{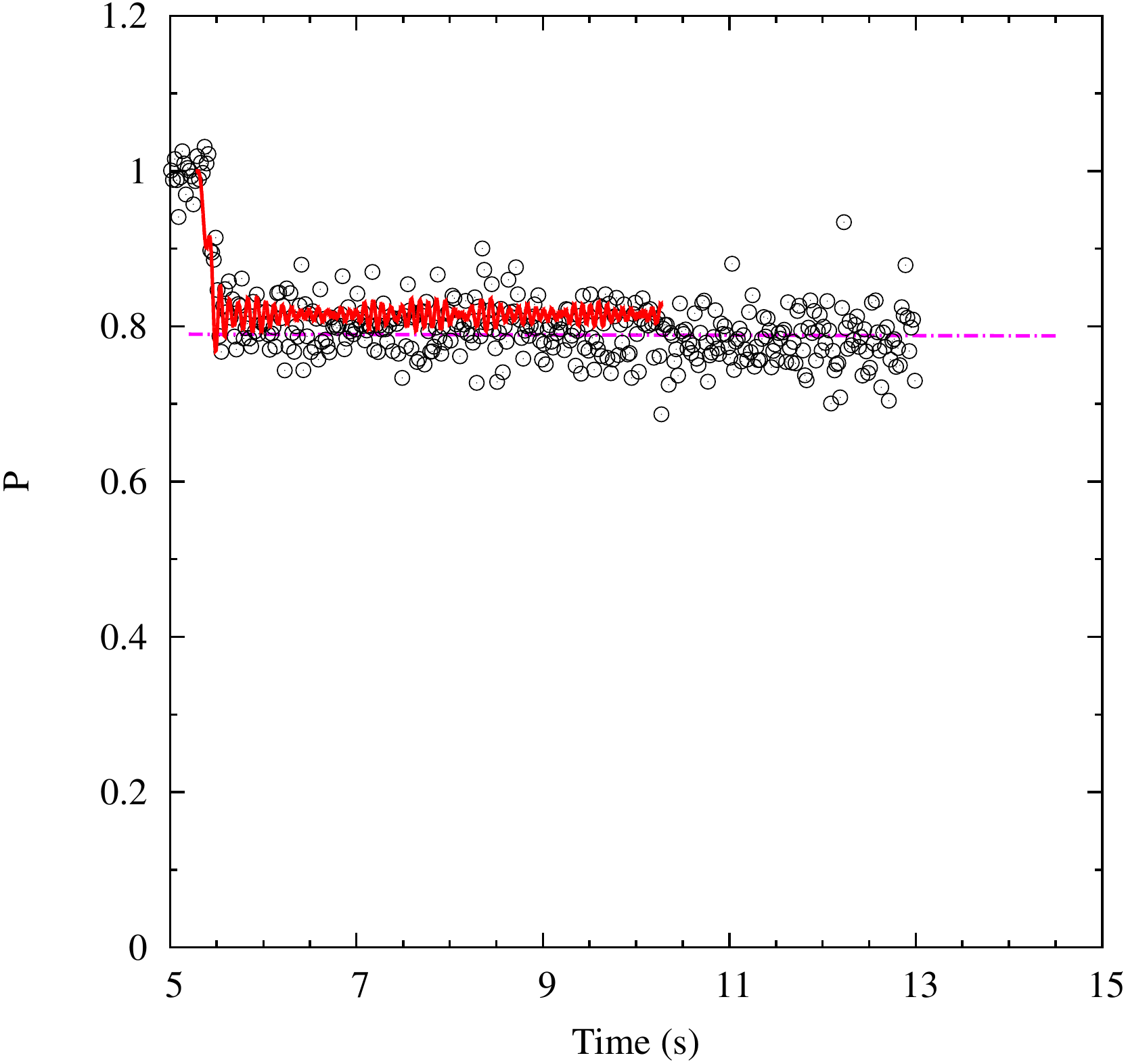}
\caption{\small
\label{fig:run85}
Plot of the data (circles) in the revised datafile sent to me by the collaboration for Run 85,
which is the data point at 871442 Hz in Fig.~22 in \cite{Benati_etal_2012}.
The solid curve is the output of a tracking simulation using a resonant frequency of 871434.4 Hz.
The dotdash line indicates a polarization of 0.78982, which is the revised value sent to me by the collaboration
(see Table \ref{tb:datafig22}).
}
\end{figure}

I display plots of the data and my simulation results in Figs.~\ref{fig:run90}--\ref{fig:run85}.
Once again, the plots are arranged in ascending order of the rf solenoid frequency, i.e.~Runs 90 and 85.
As before, the data are plotted as circles and the solid curve plots the output from my tracking simulation.
The dotdash line is the revised average polarization level sent to me by the collaboration
(see Table \ref{tb:datafig22}).
The simulation result for Run 90 (Fig.~\ref{fig:run90}) fits the data well.
The simulation result for Run 85 (Fig.~\ref{fig:run90}) is somewhat higher than the data.
I shall return to Run 85 below.

\vfill\pagebreak
\subsection{Runs 82, 83, 84 and 87: the `triangle points'}
By now I had fits for most of the data points close to the center of the resonance (except Run 87).
However, I had difficulty with Runs 83 and 84, which are the two leftmost points (lowest frequencies) in 
Fig.~22 in \cite{Benati_etal_2012},
I also had difficulty with Run 82, which is the rightmost point (highest frequency) in 
Fig.~22 in \cite{Benati_etal_2012},
These are points far from the center of the resonance, at frequencies of 817412 Hz, 817422 Hz and 871452 Hz, 
respectively about 22 Hz and 12 Hz below the center of the resonance and 18 Hz above the center of the resonance.
The polarizations are high, respectively reported as
0.93549, 0.85472 and 0.93168 in Table \ref{tb:datafig22}.
I also had difficulty fitting Run 87, at 871427 Hz.

\begin{figure}[!htb]
\centering
\includegraphics[width=0.75\textwidth]{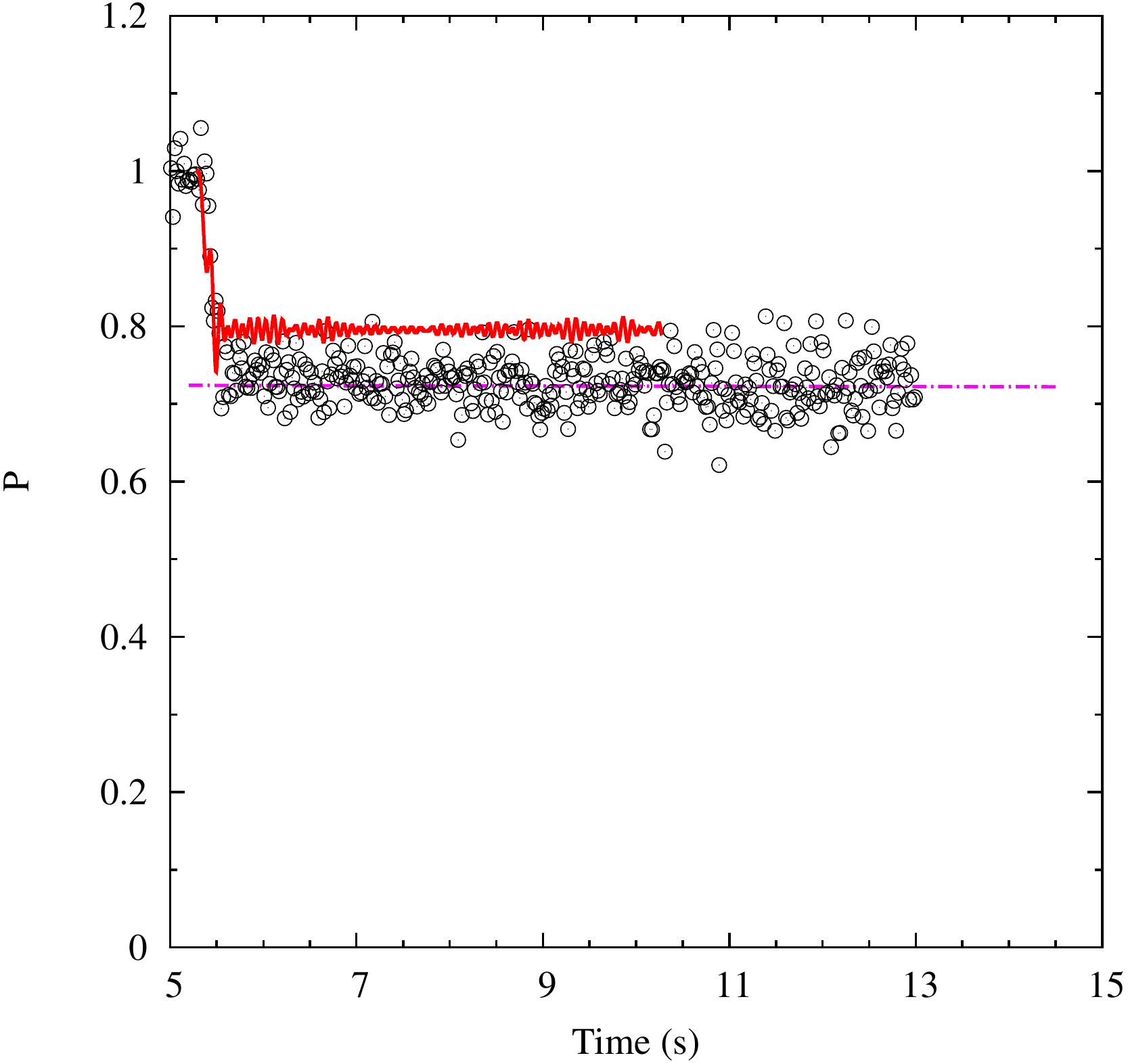}
\caption{\small
\label{fig:run87pt1}
Plot of the data (circles) in the revised datafile sent to me by the collaboration for Run 87,
which is the data point at 871427 Hz in Fig.~22 in \cite{Benati_etal_2012}.
The solid curve is the output of a tracking simulation.
The dotdash line indicates a polarization of 0.72391, which is the revised value sent to me by the collaboration
(see Table \ref{tb:datafig22}).
}
\end{figure}

\begin{figure}[!htb]
\centering
\includegraphics[width=0.75\textwidth]{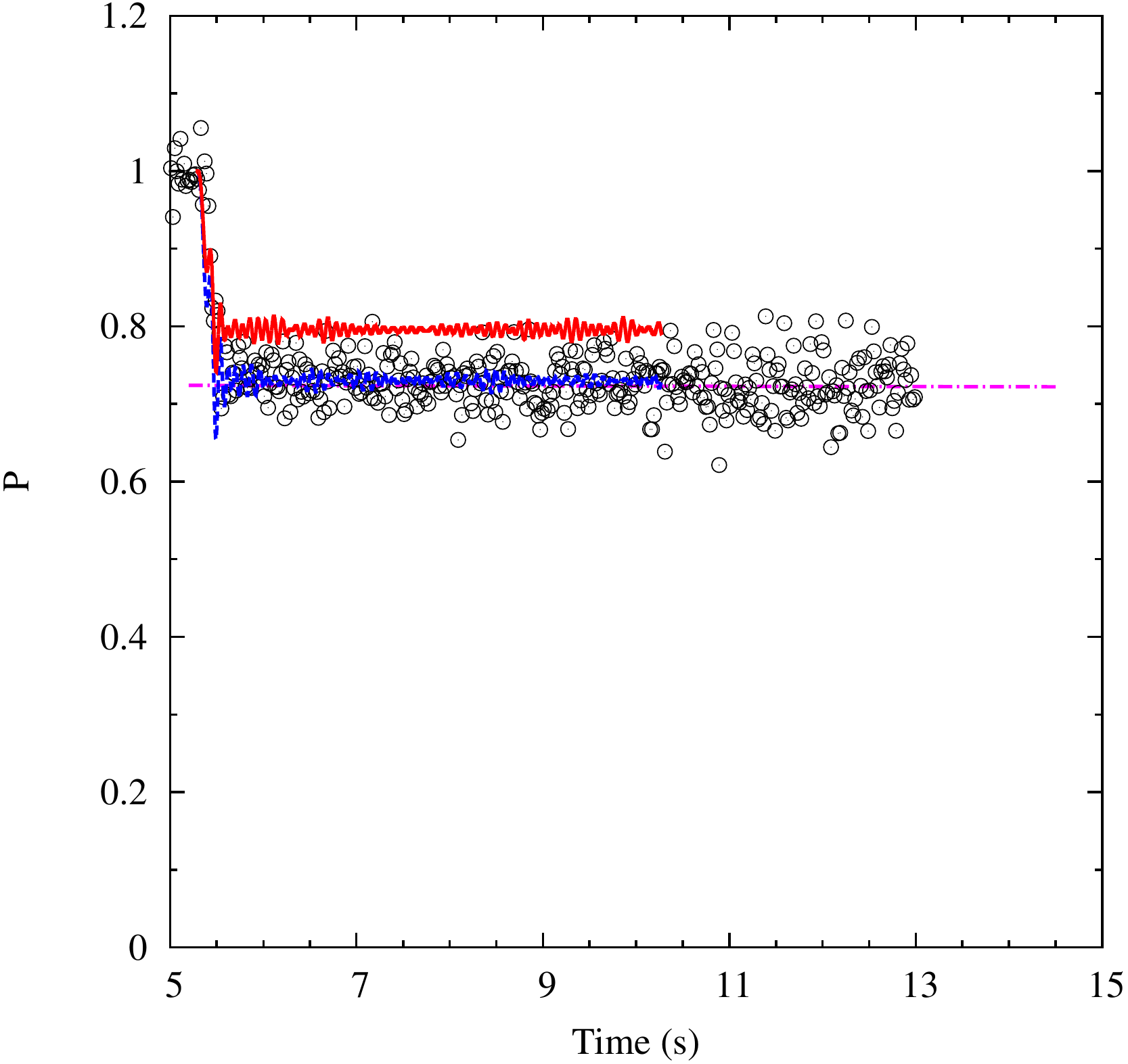}
\caption{\small
\label{fig:run87_2}
The same as Fig.~\ref{fig:run87pt1},
with an extra curve (dashed), calculated using an r.m.s.~relative momentum spread of $\sigma_p = 6.0\times 10^{-4}$.
The dotdash line again indicates the average polarization level of the data.
}
\end{figure}

I begin with Run 87.
I display a plot of the data and my simulation results in Fig.~\ref{fig:run87pt1}.
As before, the data are plotted as circles and the solid curve plots the output from my tracking simulation.
The dotdash line is the revised average polarization level of 0.72391 sent to me by the collaboration
(see Table \ref{tb:datafig22}).
It is clear that the simulation result does not match the data.
I therefore searched for alternative hypotheses which might explain the discrepancy.
As I have stated above, I initially tried a ramp time of 0.02 s.
This worked, but I was informed that the ramp time was 0.2 s for all the data points in 
Fig.~22 in \cite{Benati_etal_2012},
I ran a simulation using a smaller r.m.s.~relative momentum spread of $\sigma_p = 6.0\times 10^{-4}$.
The output is plotted as the dashed curve in Fig.~\ref{fig:run87_2},
which is otherwise the same as Fig.~\ref{fig:run87pt1}.
The dashed curve matches the data well.

\begin{figure}[!htb]
\centering
\includegraphics[width=0.75\textwidth]{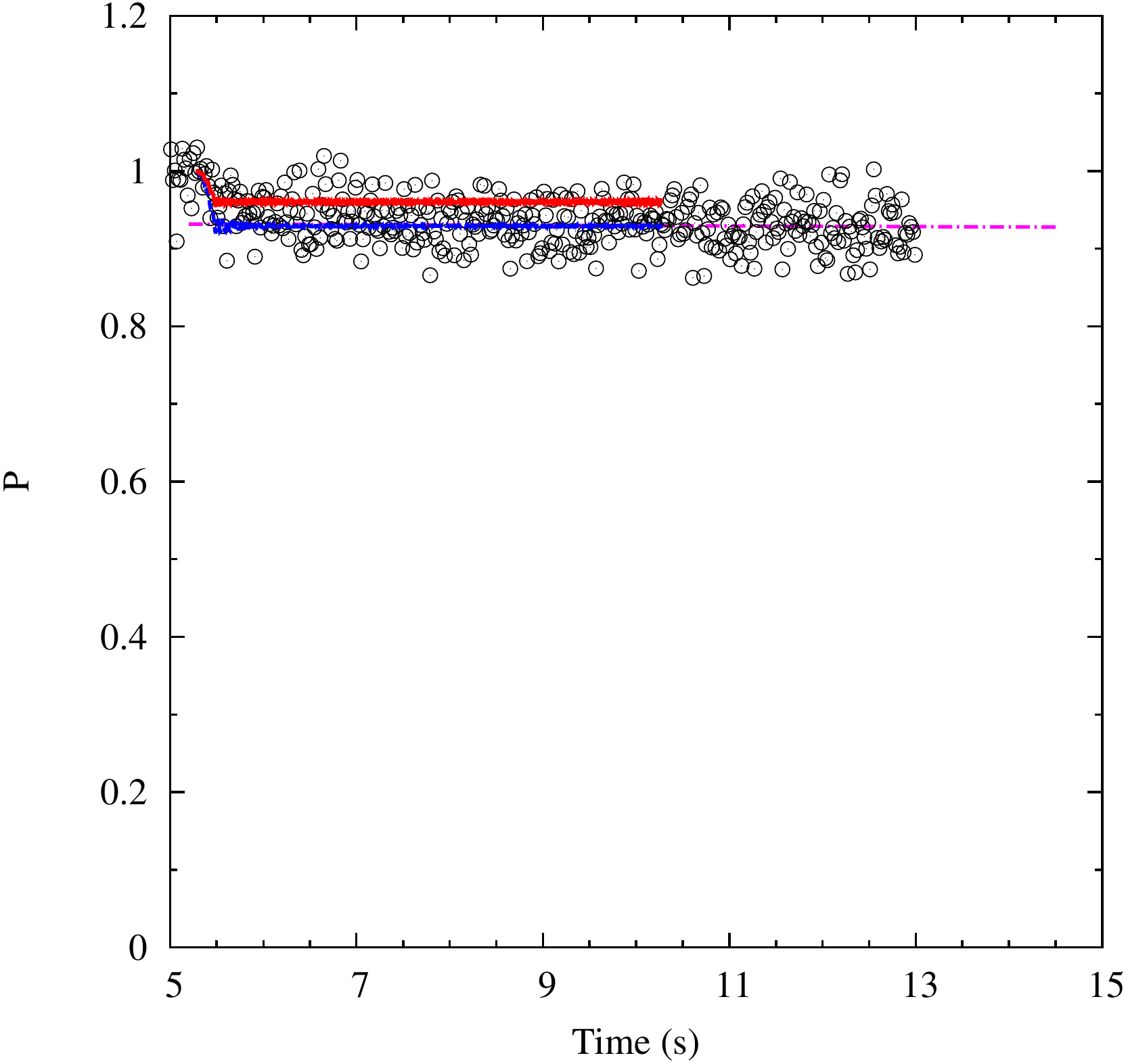}
\caption{\small
\label{fig:run82_2}
Plot of the data (circles) in the revised datafile sent to me by the collaboration for Run 82,
which is the data point at 871452 Hz in Fig.~22 in \cite{Benati_etal_2012}.
The solid and dashed curves plot the outputs of tracking simulations using 
$\sigma_p = 8.02 \times 10^{-4}$ and $\sigma_p = 6.0 \times 10^{-4}$, respectively.
The dotdash line indicates a polarization of 0.93168, which is the revised value sent to me by the collaboration
(see Table \ref{tb:datafig22}).
}
\end{figure}

\begin{figure}[!htb]
\centering
\includegraphics[width=0.75\textwidth]{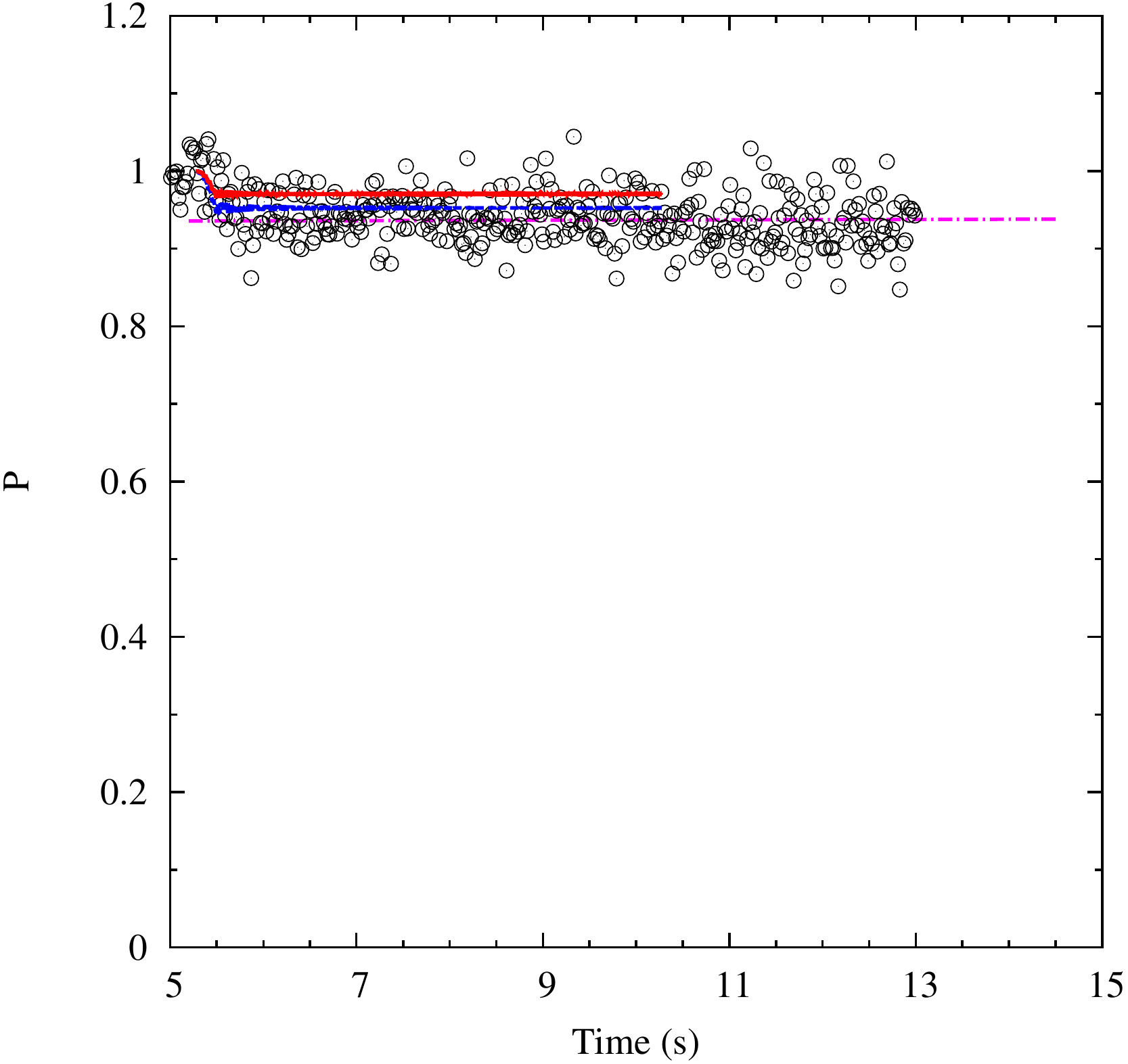}
\caption{\small
\label{fig:run83_2}
Plot of the data (circles) in the revised datafile sent to me by the collaboration for Run 83,
which is the data point at 871412 Hz in Fig.~22 in \cite{Benati_etal_2012}.
The solid and dashed curves plot the outputs of tracking simulations using 
$\sigma_p = 8.02 \times 10^{-4}$ and $\sigma_p = 6.0 \times 10^{-4}$, respectively.
The dotdash line indicates a polarization of 0.93549, which is the revised value sent to me by the collaboration
(see Table \ref{tb:datafig22}).
}
\end{figure}

\begin{figure}[!htb]
\centering
\includegraphics[width=0.75\textwidth]{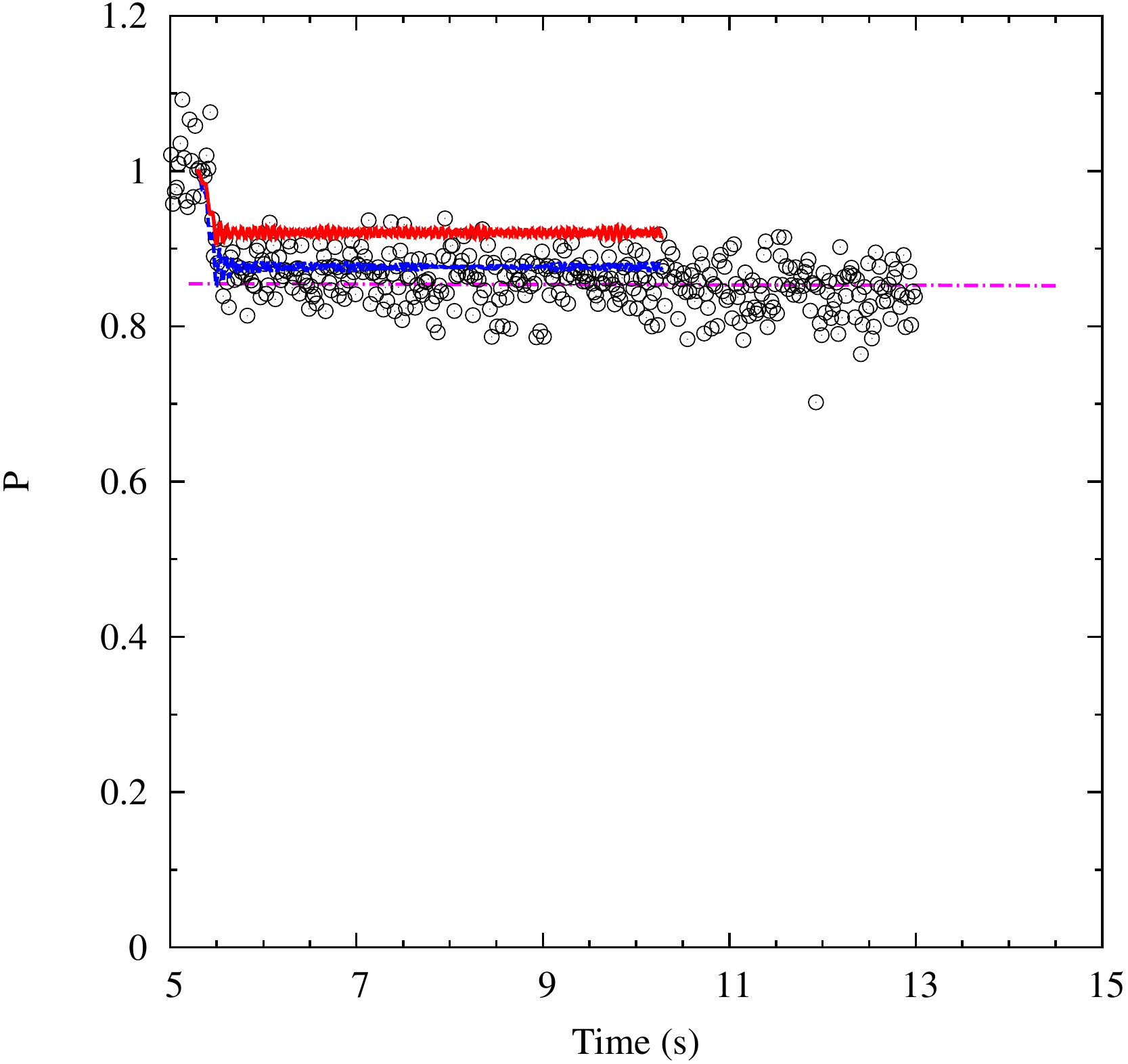}
\caption{\small
\label{fig:run84_2}
Plot of the data (circles) in the revised datafile sent to me by the collaboration for Run 84,
which is the data point at 871422 Hz in Fig.~22 in \cite{Benati_etal_2012}.
The solid and dashed curves plot the outputs of tracking simulations using 
$\sigma_p = 8.02 \times 10^{-4}$ and $\sigma_p = 6.0 \times 10^{-4}$, respectively.
The dotdash line indicates a polarization of 0.85472, which is the revised value sent to me by the collaboration
(see Table \ref{tb:datafig22}).
}
\end{figure}

To save time, let me state here that the use of $\sigma_p = 6.0\times 10^{-4}$
gave good fits to the data for all of Runs 82, 83 and 84.
I plot the data and fits in Figs.~\ref{fig:run82_2}--\ref{fig:run84_2}, for Runs 82, 83 and 84, respectively.
As before, the data are plotted as circles.
The solid and dashed curves plot the outputs from my tracking simulations using
$\sigma_p = 8.02\times 10^{-4}$ and $\sigma_p = 6.0\times 10^{-4}$, respectively.
The dotdash line indicates the revised average polarization level sent to me by the collaboration
(see Table \ref{tb:datafig22}).

{\em Caveat:}
I employed a resonant frequency of 871434.0 Hz for all of the simulations in this section.
I found that a resonant frequency of 817434.4 Hz did not work for Run 87.
For the other cases, the frequencies were so far from the center of the resonance that
small changes to the resonant frequency had a negligible effect on the simulation results.
I did not consider it reasonable to shift the resonant frequency by several Hz.
I therefore employed a resonant frequency of 871434.0 Hz in all of 
Figs.~\ref{fig:run87pt1}--\ref{fig:run84_2}.

\vfill\pagebreak
\subsection{Run 85 revisited}
\begin{figure}[!htb]
\centering
\includegraphics[width=0.75\textwidth]{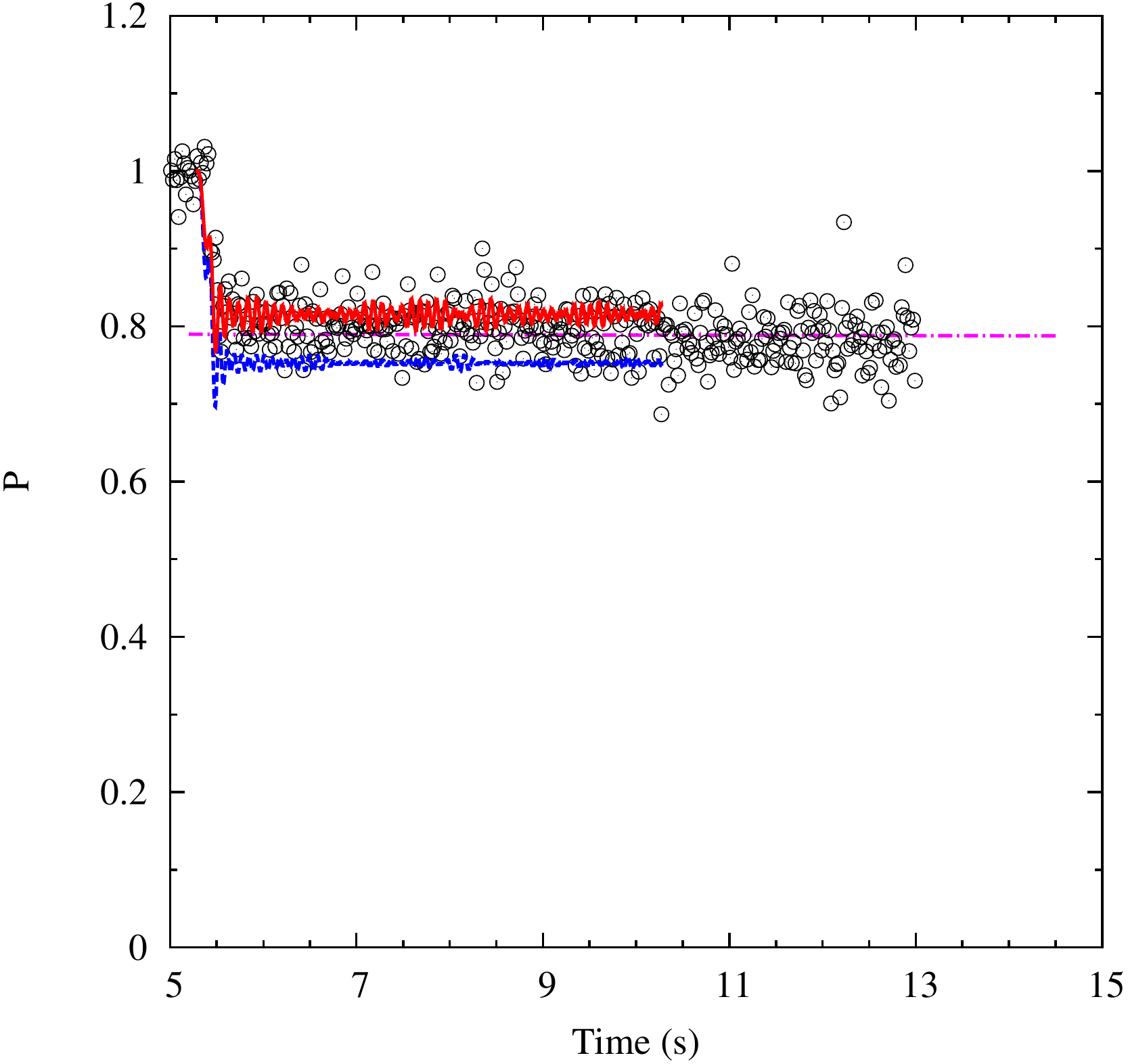}
\caption{\small
\label{fig:run85_3}
The same as Fig.~\ref{fig:run85},
with an extra curve (dashed), calculated using an r.m.s.~relative momentum spread of $\sigma_p = 6.0\times 10^{-4}$.
The dotdash line again indicates the average polarization level of the data.
}
\end{figure}

I remarked earlier that the simulation result for Run 85, shown in Fig.~\ref{fig:run85}, was somewhat higher than the data.
I therefore ran a second simulation using a smaller r.m.s.~relative momentum spread of $\sigma_p = 6.0\times 10^{-4}$.
The result is shown as the dashed curve in Fig.~\ref{fig:run85_3}.
This time the simulation result is too low.

\begin{figure}[!htb]
\centering
\includegraphics[width=0.75\textwidth]{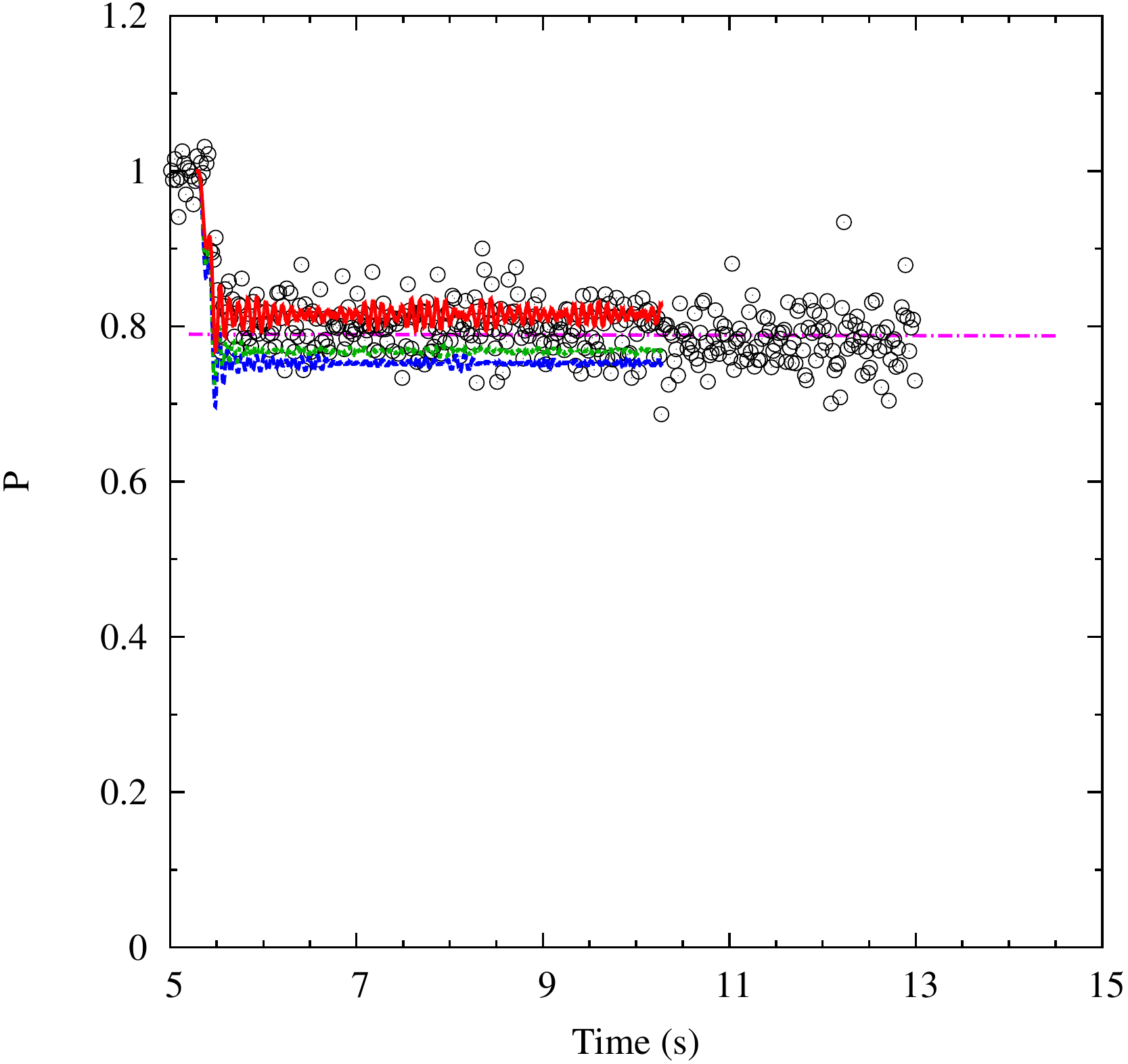}
\caption{\small
\label{fig:run85_4}
The same as Fig.~\ref{fig:run85_3},
with an extra curve (dotted), calculated using an r.m.s.~relative momentum spread of $\sigma_p = 6.0\times 10^{-4}$
using a resonant frequency of 817434.0 Hz.
All the other curves employ a resonant frequency of 817434.4 Hz.
The dotdash line again indicates the average polarization level of the data.
}
\end{figure}

Note that this point is at a frequency of 871442 Hz, i.e.~above the resonance center.
Recall this point was fitted using a resonant frequency of 817434.4 Hz.
Hence lowering the resonant frequency to 817434.0 Hz will raise the polarzation level in a tracking simulation.
I therefore ran a third simulation using $\sigma_p = 6.0\times 10^{-4}$
and a resonant frequency of 817434.0 Hz.
The result is shown as the dotted curve in Fig.~\ref{fig:run85_4}.
The result is higher than the dashed curve, as expected, but it is still too low.

Hence I conclude that I have no simple way to fit the data in Run 85
to the same quality which I can achieve for most of the other data points.
Obviously, I could retrofit a value of $\sigma_p$ to fit the data, 
but I have explained that such retrofitting has no predictive value.
I therefore conclude, as I stated above, that I have is no simple way to fit the data in Run 85.

\vfill\pagebreak
\section{\label{sec:syncosc} Reduction of resonance width by synchrotron oscillations}
\begin{figure}[!htb]
\centering
\includegraphics[width=0.75\textwidth]{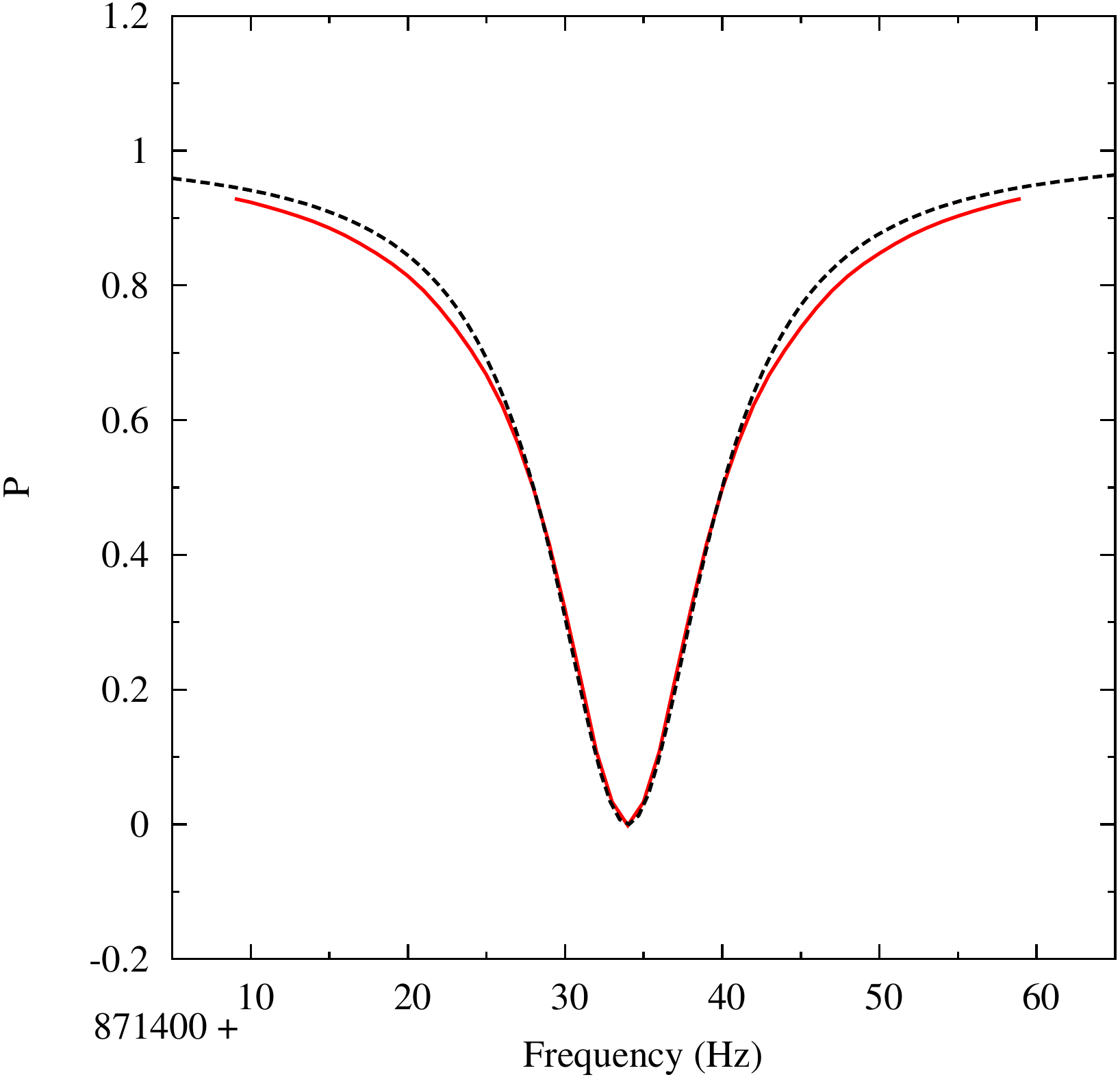}
\caption{\small
\label{fig:sweep-uncooled_no_oscs}
Plot of simulations results for a monochromatic beam, 
using a ramp time of 0.2 s and a resonant frequency of 871434.0 Hz.
One particle on the reference orbit was tracked.
The solid curve displays the simulation result and the dotdash curve is a Lorentzian fit.
}
\end{figure}

It is well known that the synchrotron oscillations induce `satellite' sideband resonances
and reduce the width of the parent resonance.
Let us estimate the magnitude of this reduction.
I derived an analytical expression for the reduction of the width of the parent resonance due to synchrotron oscillations in
\cite{ManeSyncTunMod2013}.
I derived a reduction factor of $e^{-\xi^2/2}\sqrt{I_0(\xi^2)}$, where the parameter $\xi$ is defined in
\cite{ManeSyncTunMod2013}.
For the present experiment, 
$\xi \simeq -0.169 \times 10^4 \times \sigma_p$.
Hence 
\bq
\xi \simeq \begin{cases} -1.356 & \qquad (\sigma_p = 8.02 \times 10^{-4}) \,, 
\\
-1.015 & \qquad (\sigma_p = 6.0 \times 10^{-4}) \,.
\end{cases}
\eq
This yields reduction factors of
\bq
e^{-\xi^2/2}\sqrt{I_0(\xi^2)} \simeq \begin{cases} 0.57 & \qquad (\sigma_p = 8.02 \times 10^{-4}) \,, 
\\
0.67 & \qquad (\sigma_p = 6.0 \times 10^{-4}) \,.
\end{cases}
\eq
For the tracking studies, I simulated the parent resonance by tracking one particle on the reference orbit,
with a ramp time of 0.2 s and a resonant frequency of 871434.0 Hz.
The simulation results are plotted in 
Fig.~\ref{fig:sweep-uncooled_no_oscs}.
The solid curve displays the simulation result and the dotdash curve is a Lorentzian fit.
We see that close to the center of the resonance the simulation result is approximately a Lorentzian
but it deviates farther away from the center.
The FWHM resonance width is 12.0 Hz, corresponding to an effective resonance strength\footnote{
In my initial analysis of the resonance dip of the uncooled beam in
Fig.~22 in \cite{Benati_etal_2012},
I compared the resonance width to the `na{\"\i}ve' resonance strength of $\varepsilon_{\rm FWHM} = 2.66 \times 10^{-5}$,
but this did not take into account the effects of the initial ramp.
}
of $\varepsilon_{\rm FWHM} = 1.6 \times 10^{-4}$.
From Fig.~\ref{fig:sweep-uncooled-2}, although the curves are not exactly Lorentzians,
we can determine the FWHM resonance widths for the two cases 
(the two r.m.s.~relative momentum spreads)
as 5.4 Hz and 7.8 Hz, respectively. This yields (`$w$' for width)
\bq
\frac{w_{\rm fit}}{w_{\rm parent}} \simeq \begin{cases} 
\displaystyle \frac{5.4}{12.0} \simeq 0.45 & \qquad (\sigma_p = 8.02 \times 10^{-4}) \,, \phantom{\Biggl|} 
\\
\displaystyle \frac{7.8}{12.0} \simeq 0.67 & \qquad (\sigma_p = 6.0 \times 10^{-4}) \,. \phantom{\Biggl|} 
\end{cases}
\eq
There is an approximate agreement with the analytical theory (better for $\sigma_p = 6.0 \times 10^{-4}$).
It would be desirable to measure resonance dips for both the parent and sideband resonances, 
all under the same experimental conditions.

\vfill\pagebreak
\section{\label{sec:fig21171216} Fits for other data for the uncooled beam in \cite{Benati_etal_2012}}
\subsection{Fig.~21 in \cite{Benati_etal_2012}}

\begin{figure}[!htb]
\centering
\includegraphics[width=0.75\textwidth]{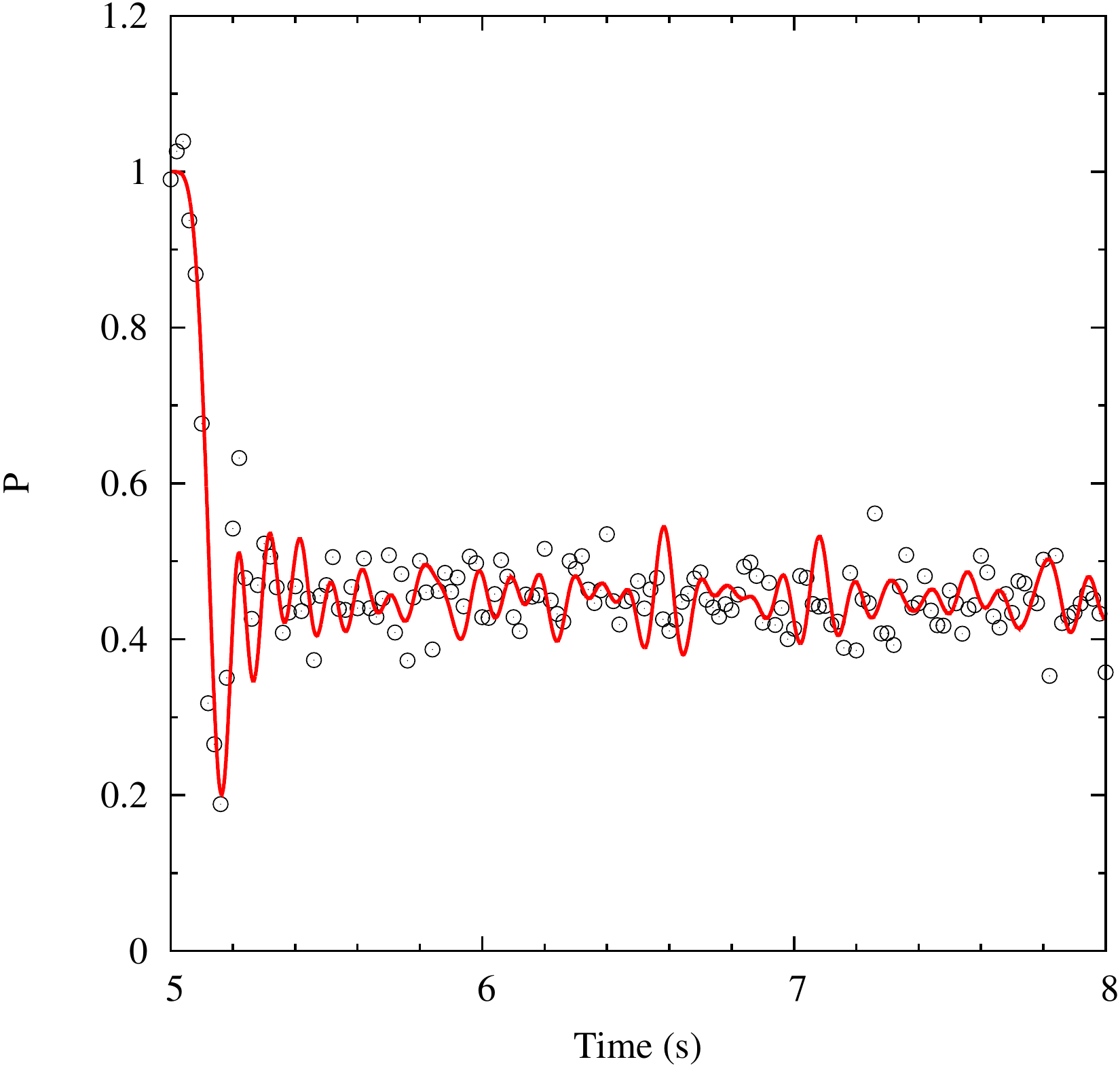}
\caption{\small
\label{fig:fig21}
Plot of the data (circles) in the revised datafile sent to me by the collaboration for Fig.~21 in \cite{Benati_etal_2012}
(which was Run 86).
The solid line shows the output from the tracking simulation.
The data and simulation are actually a blown up version of the initial portion of Fig.~\ref{fig:run86}.
}
\end{figure}

In my recent paper \cite{ManeSyncTunMod2013}, I claimed that the data in Fig.~21 in \cite{Benati_etal_2012} was 4 Hz off resonance.
{\em This is incorrect.}
The data in Fig.~21 in \cite{Benati_etal_2012} were taken at an rf solenoid frequency of 817432 Hz,
and were Run 86 (see in Table \ref{tb:datafig22}).
I now know that the resonance center for this run was at 817434.4 Hz,
hence the data were 2.4 Hz below resonance.
I display the (revised) data (as circles) and tracking output (solid curve) in Fig.~\ref{fig:fig21}.
The data and simulation are actually a blown up version of the initial portion of Fig.~\ref{fig:run86}.

\vfill\pagebreak
\subsection{Fig.~17 in \cite{Benati_etal_2012}}

\begin{figure}[!htb]
\centering
\includegraphics[width=0.75\textwidth]{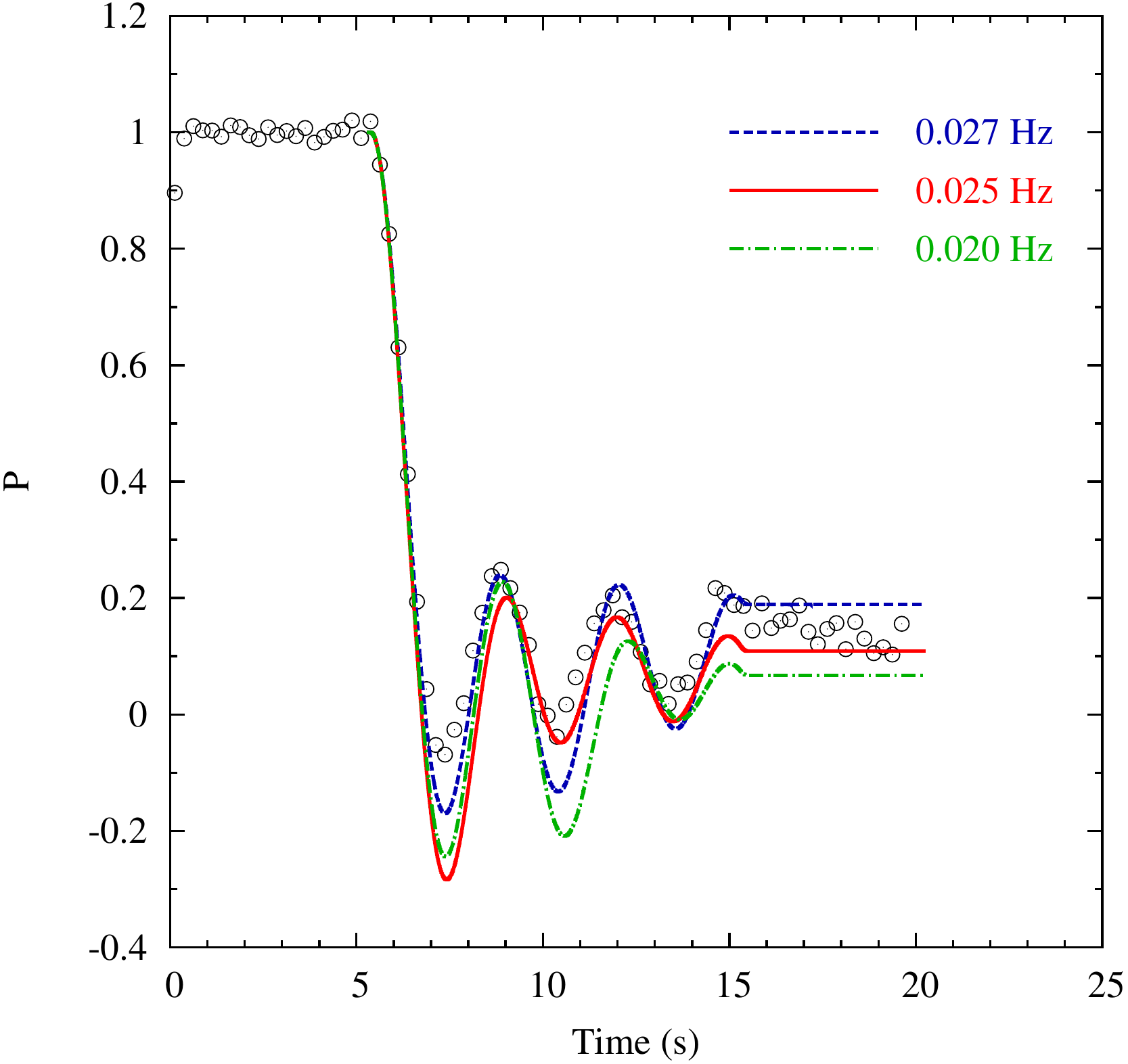}
\caption{\small
\label{fig:track_fig17}
Data points in Fig.~17 in \cite{Benati_etal_2012} (circles).
The tracking simulations used a resonance strength of $\varepsilon_{\rm FWHM} = 8.87 \times 10^{-7}$.
The rf solenoid frequency was displaced from the resonance center by 
$0.027$ Hz (dashed line),
$0.025$ Hz (solid line)
and by $0.02$ Hz (dotdashed line).
The rf solenoid was ramped down to zero at $t=15$ s and the simulation results are flat after that.
The dashed and solid curves (offsets $0.027$ and $0.025$ Hz, respectively) seem to yield the best fit the data.
All three curves seem to go down too low at the minima.
All three curves seem to reproduce the salient features of the data, including a nonzero average of about $P\simeq0.1$ at the end.
}
\end{figure}

I offered an alternative hypothesis for the data in Fig.~17 in \cite{Benati_etal_2012} 
in my recent paper \cite{ManeSyncTunMod2013}.
Here I present fresh tracking simulations, following the experimental procedure of the data acquisition (ramp up time, etc.).
I display the data, together with tracking results for three choices of frequency offsets in 
Fig.~\ref{fig:track_fig17}.\footnote{The revision of the normalization of the data for the uncooled resonance curve 
did not extend to the data from Fig.~17 in \cite{Benati_etal_2012}.
That data remains as it is in \cite{Benati_etal_2012}.}
I was also informed that the rf solenoid was ramped down to zero at $t=15$ s,
hence my simulation results are flat after that time.
My inference is the same as before, that the data in Fig.~17 in \cite{Benati_etal_2012} 
can be explained by assuming that the run was made 0.025 Hz off resonance.
For brevity, I simply repeat the statements I made in \cite{ManeSyncTunMod2013}:
\begin{quote} 
Next, Fig.~17 in \cite{Benati_etal_2012} is interesting.
The theory curve by the authors in Fig.~17 oscillates and decays with an average value of zero.
However, the data appear to oscillate and average to a nonzero value of approximately $0.1$.
The authors state that 
``Figures 16 and 17 show time curves calculated with the
model of Fig.~12, but with different solenoid strengths. In
Fig.~17, agreement is not as good. In particular, more tracks
are needed with small amplitudes and faster oscillation
frequencies to better match the first two oscillations in
the polarization. Adding such tracks improves the agreement
for that feature, but this change overestimates the
oscillations at larger times.''
Later the authors state
``During the
several hours spent taking these measurements, it is likely
that the distribution of particles within the rf bucket
changed, even if the bucket potential is unchanged.''
Note that if the distribution of particles within the rf bucket changed,
such a change would be independent of the rf solenoid and would, in principle, 
affect all of the studies reported in \cite{Benati_etal_2012}.
 
I offer an alternative explanation for the data in Fig.~17 in \cite{Benati_etal_2012}.
The data in Fig.~17 were taken with a very small rf solenoid strength ($\varepsilon_{\rm FWHM} = 8.87 \times 10^{-7}$)
and an uncooled beam.
I scaled down the resonance width from Fig.~22 in \cite{Benati_etal_2012},
which was an uncooled beam and $\varepsilon_{\rm FWHM} = 2.66\times 10^{-6}$ and FWHM resonance width of $7.5$ Hz,
to deduce that the FWHM resonance width for the data in Fig.~17 in \cite{Benati_etal_2012} 
was approximately 30 times smaller, i.e.~only $0.25$ Hz.
Such a small resonance width is within the uncertainty for the exact location of the resonance center.
Hence I hypothesize that the rf solenoid frequency in Fig.~17 in \cite{Benati_etal_2012}
was {\em not} exactly on resonance.
The results of my simulations \dots 
{\em [At this stage, I present fresh simulations, see Fig.~\ref{fig:track_fig17}].}
\end{quote}

\vfill\pagebreak
\subsection{Figs.~12 and 16 in \cite{Benati_etal_2012}}

\begin{figure}[!htb]
\centering
\includegraphics[width=0.75\textwidth]{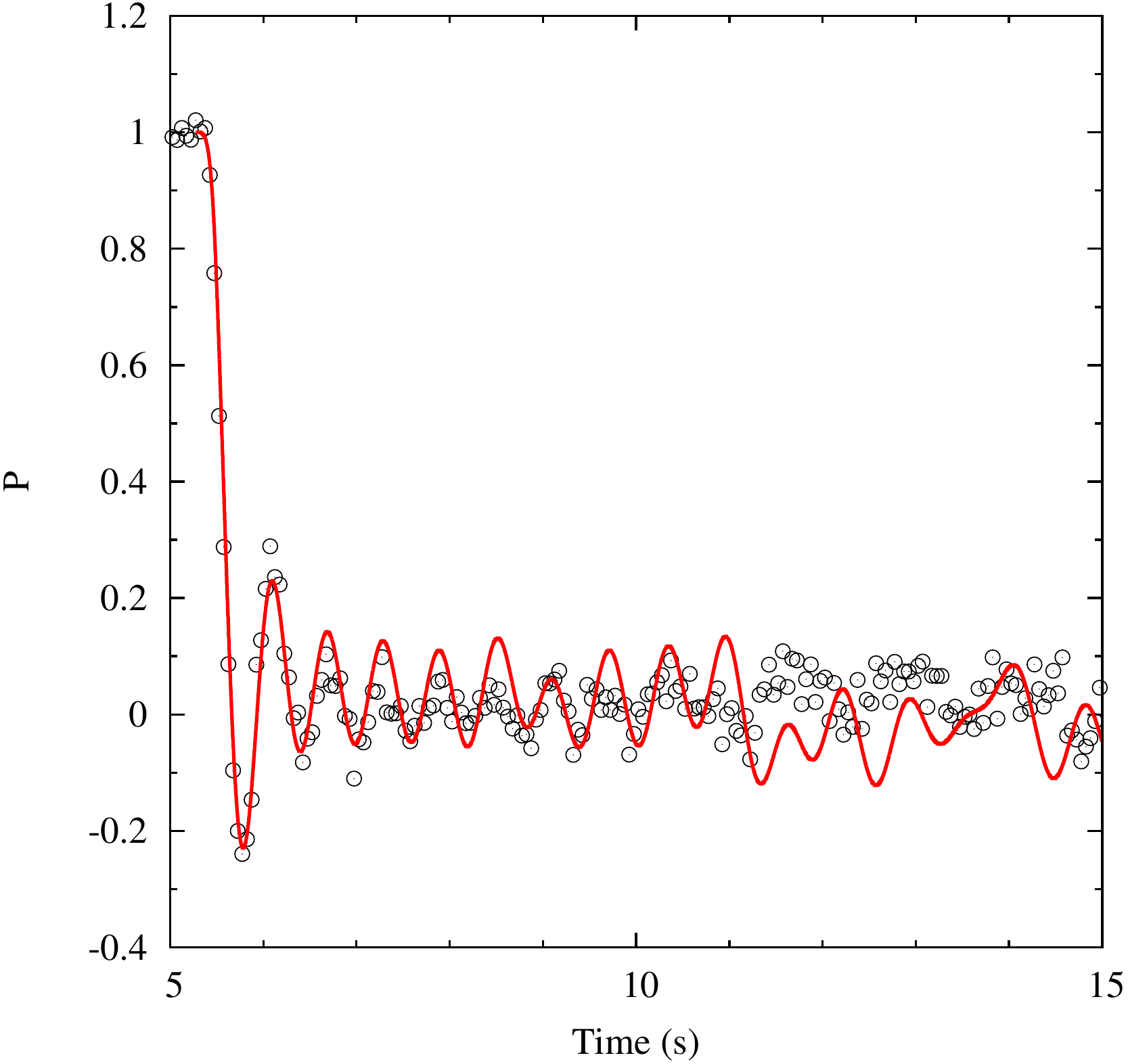}
\caption{\small
\label{fig:track_fig12}
Data points in Fig.~12 in \cite{Benati_etal_2012} (circles).
These were measurements made on resonance (817434.0 Hz) with an uncooled beam and
a resonance strength of $\varepsilon_{\rm FWHM} = 4.43 \times 10^{-6}$.
The tracking simulation results are shown as the solid curve.
}
\end{figure}

\begin{figure}[!htb]
\centering
\includegraphics[width=0.75\textwidth]{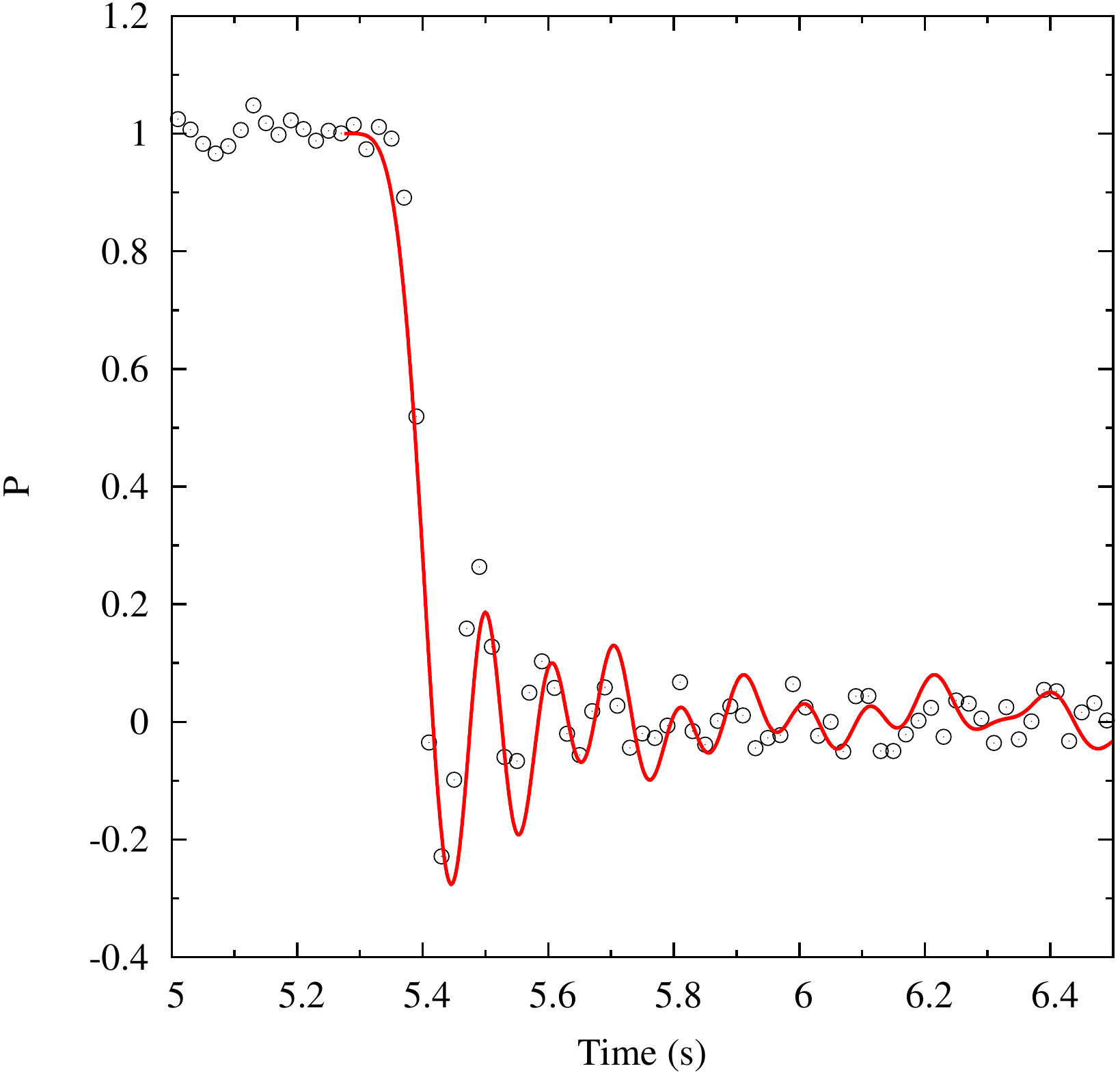}
\caption{\small
\label{fig:track_fig16a}
Data points in Fig.~16 in \cite{Benati_etal_2012} (circles).
These were measurements made on resonance (817434.0 Hz) with an uncooled beam and
a resonance strength of $\varepsilon_{\rm FWHM} = 2.66 \times 10^{-5}$.
The data are plotted in the range $5 \le t \le 6.5$ s, as in \cite{Benati_etal_2012}. 
The tracking simulation results are shown as the solid curve.
}
\end{figure}

\begin{figure}[!htb]
\centering
\includegraphics[width=0.75\textwidth]{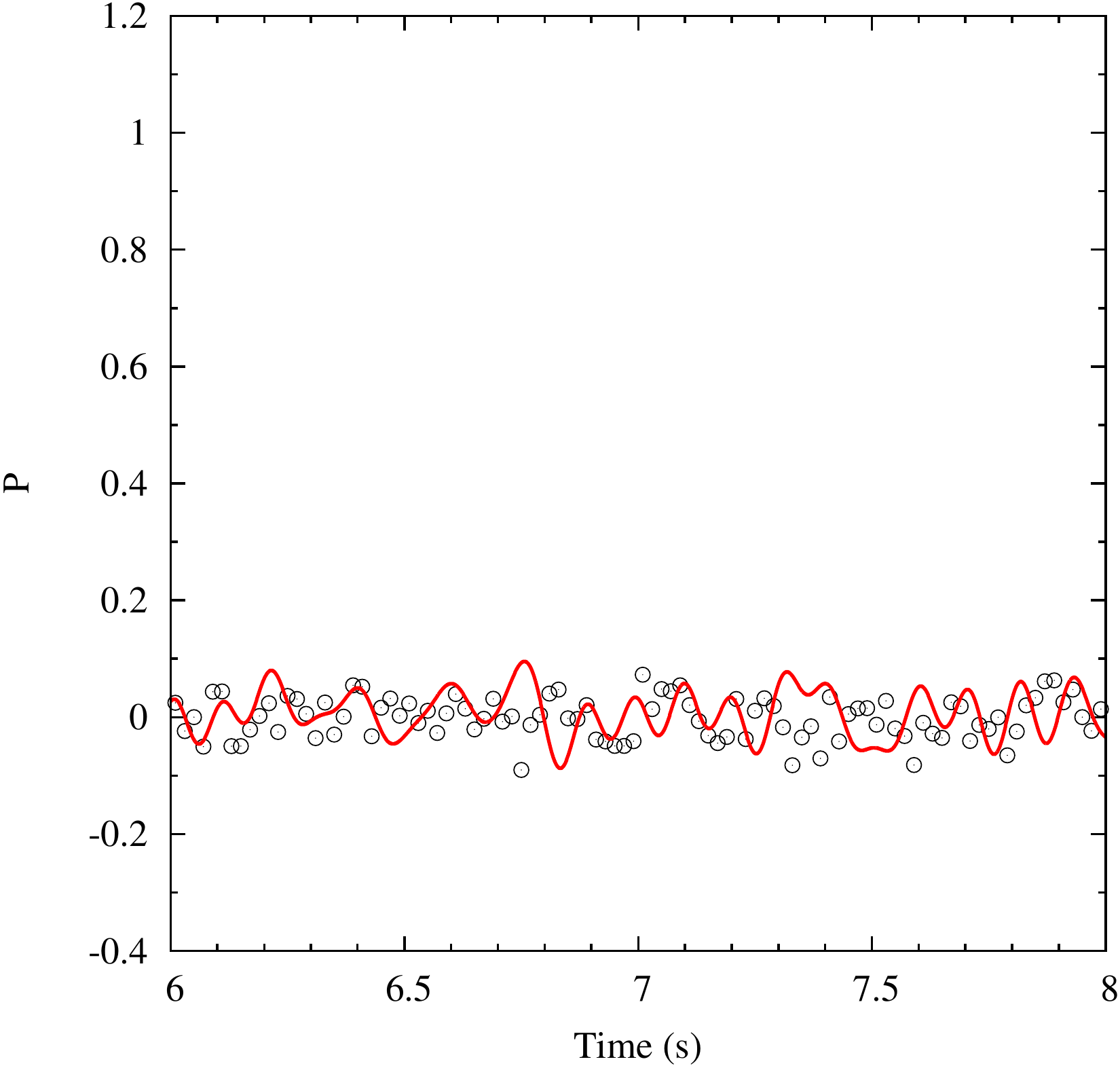}
\caption{\small
\label{fig:track_fig16b}
The circles plot the continuation of the points in Fig.~\ref{fig:track_fig16a},
i.e.~Fig.~16 in \cite{Benati_etal_2012}.
These were measurements made on resonance (817434.0 Hz) with an uncooled beam and
a resonance strength of $\varepsilon_{\rm FWHM} = 2.66 \times 10^{-5}$,
but plotted in the range $6 \le t \le 8$ s, which was not actually displayed in \cite{Benati_etal_2012}. 
The tracking simulation results are shown as the solid curve.
}
\end{figure}

The data in Figs.~12 and 16 in \cite{Benati_etal_2012} 
were for measurements made on resonance (817434.0 Hz) with an uncooled beam and a resonance strength of 
$\varepsilon_{\rm FWHM} = 4.43 \times 10^{-6}$ and $2.66 \times 10^{-5}$, respectively.
I plot the data and display fits using my tracking simulations in 
Figs.~\ref{fig:track_fig12} and \ref{fig:track_fig16a}, respectively.
The data shown in these plots are the same as those in \cite{Benati_etal_2012}, i.e.~not revised.
The data are plotted as circles and my simulation results are shown as the solid curves.
In Fig.~\ref{fig:track_fig16a}, I plot 
the data in the range $5 \le t \le 6.5$ s, as in \cite{Benati_etal_2012},
while in Fig.~\ref{fig:track_fig16b}, I continue the plot 
into the range $6 \le t \le 8$ s, which was not actually displayed in \cite{Benati_etal_2012}. 
My principal conclusion is that a tracking simulation using a Gaussian distribution of the particle orbits
is able to reproduce the salient features of the data.

\vfill\pagebreak
\section{\label{sec:conc} Conclusion}
Most of my analysis was to fit the data for the resonance dip in Fig.~22 in \cite{Benati_etal_2012}.
To do so, it was essential to follow the experimental procedure for the data acquisition.
I showed that the points cannot all be fitted by a single resonant frequency,
and that some of the outlying points do not lie on the same resonance curves as those closer to the center.
I also argued that the data in Fig.~22 in \cite{Benati_etal_2012}
are not sufficiently precise to definitively support any specific distribution of the particle orbits:
I could in principle retrofit every point using a Gaussian distribution,
but such a fit would have no predictive power or explanatory value.

I also corrected an error in \cite{ManeSyncTunMod2013} for the frequency offset from the resonance center
for the data in Fig.~21 in \cite{Benati_etal_2012};
it is 2.4 Hz, and not 4 Hz as I claimed in \cite{ManeSyncTunMod2013}.
To see this, it was essential to follow the experimental procedure and include an initial ramp for the amplitude of the rf solenoid.
I also had to determine that the resonant frequency for this run was 817434.4 Hz and not 817434.0 Hz, as I explained above.

I also presented more detailed fits for the data in Fig.~17 in \cite{Benati_etal_2012},
following the experimental procedure for the data acquisition.
It was stated in \cite{Benati_etal_2012} that these measurements were made on resonance with
a weak resonance strength of $\varepsilon_{\rm FWHM} = 8.87 \times 10^{-7}$.
I suggested in \cite{ManeSyncTunMod2013} 
that the nonzero average of the oscillations of the polarization in Fig.~17 in \cite{Benati_etal_2012}
is consistent with an interpretation that the measurements were actually a fraction of a Hz off resonance.
I reiterate the above claim, using more detailed simulations.
As I stated in \cite{ManeSyncTunMod2013}, I suggest the measurements were 0.025 Hz off resonance.

I also displayed the results of tracking simulations to fit the data in Figs.~12 and 16 in \cite{Benati_etal_2012},
which were for measurements made on resonance with an uncooled beam and a resonance strength of 
$\varepsilon_{\rm FWHM} = 4.43 \times 10^{-6}$ and $2.66 \times 10^{-5}$, respectively.
I demonstrated that a tracking simulation using a Gaussian distribution of the particle orbits
was able to reproduce the salient features of the data in both cases.

I thank the collaboration again for sending me files of their data and 
for explaining the data acquisition procedure employed in \cite{Benati_etal_2012}.
The authors claim in \cite{Benati_etal_2012} that their data and simulations (using non-Gaussian orbital distributions)
provide a strong test of the ``no lattice'' model they use in \cite{Benati_etal_2012}.
I see no evidence in the data to support any specific model of the particle oscillation amplitudes.
I can explain the data for the uncooled beam,
in Figs.~12, 16, 17, 21 and 22 in \cite{Benati_etal_2012},
using a Gaussian distribution of the particle orbits.

\vfill\pagebreak

\vfill\pagebreak

\begin{table}[!htb]
\centering
\begin{tabular}[width=0.75\textwidth]{rcllllrc}
\hline
\quad & Freq.~(Hz) & \quad & $\quad P$ & \quad & $\quad \Delta P$ & \quad & Run \# \\
\hline
\hline
& 871412.0    &&   0.93549 &&   0.0018270 && 83  \\
& 871422.0    &&   0.85472 &&   0.0018532 && 84  \\
& 871327.0    &&   0.72391 &&   0.0017741 && 87  \\
& 871432.0    &&   0.43449 &&   0.0018619 && 86  \\
& 871434.0    &&   0.0036147  && 0.0019661 && 91  \\
& 871435.0    &&   0.11855    && 0.0018988 && 88  \\
& 871436.0    &&   0.28409    && 0.0019772 && 89  \\
& 871437.0    &&   0.52834    && 0.0030428 && 93  \\
& 871439.0    &&   0.64286    && 0.0020189 && 90  \\
& 871442.0    &&   0.78982    && 0.0018351 && 85  \\
& 871452.0    &&   0.93168    && 0.0017655 && 82  
\\
\hline
\end{tabular}
\caption{\small
\label{tb:datafig22}
Frequency and polarization values published in Fig.~22 in \cite{Benati_etal_2012}.
The run numbers of the data acquisition are also tabulated.
Note that there was no Run 92.
}
\end{table}

\vfill\pagebreak

\begin{table}[!htb]
\centering
\begin{tabular}[width=0.75\textwidth]{rcllrcrcrc}
\hline
\quad & Freq.~(Hz) & \quad & $\quad P$ & \quad & Run \# & \quad & $f_{\rm res}$ (Hz) & \quad & $\sigma_p$ \\
\hline
& 871412.0    &&   0.93549 &&   83  && && $6.0 \times 10^{-4}$ \\
& 871422.0    &&   0.85472 &&   84  && && $6.0 \times 10^{-4}$ \\
& 871327.0    &&   0.72391 &&   87  && && $6.0 \times 10^{-4}$ \\
& 871432.0    &&   0.43449 &&   86  && 871434.4 \\
& 871434.0    &&   0.0036147  && 91  \\
& 871435.0    &&   0.11855    && 88  && 871434.4 \\
& 871436.0    &&   0.28409    && 89  && 871434.4 \\
& 871437.0    &&   0.52834    && 93  \\
& 871439.0    &&   0.64286    && 90  && 871434.4 \\
& 871442.0    &&   0.78982    && 85  && 871434.4 \\
& 871452.0    &&   0.93168    && 82  && && $6.0 \times 10^{-4}$
\\
\hline
\end{tabular}
\caption{\small
\label{tb:datafig22comment}
The data in Table \ref{tb:datafig22} with columns to indicate the resonance center and r.m.s.~relative momentum spread.
A blank resonance center means 871434.0 Hz and a blank r.m.s.~relative momentum spread means $8.02 \times 10^{-4}$.
}
\end{table}

\vfill\pagebreak

\begin{table}[!htb]
\centering
\begin{tabular}[width=0.75\textwidth]{rcrcllrcrc}
\hline
\quad & Effective Freq.~(Hz) & \quad & \quad & \quad & $\quad P$ & \quad & Run \# & \quad & Symbol \\
\hline
\hline
& 871412.0 & --  &   &&   0.93549     && 83  && Triangle \\
& 871422.0 & --  &   &&   0.85472     && 84  && Triangle \\
& 871327.0 & --  &   &&   0.72391     && 87  && Triangle \\
& 871431.6 & *   &   &&   0.43449     && 86  && Square \\
& 871434.0 &     &   &&   0.0036147    && 91  && Circle \\
& 871434.6 & *   &   &&   0.11855     && 88  && Square \\
& 871435.6 & *   &   &&   0.28409     && 89  && Square \\
& 871437.0 &     &   &&   0.52834     && 93  && Circle \\
& 871438.6 & *   &   &&   0.64286     && 90  && Square \\
& 871441.6 & *   &   &&   0.78982     && 85  && Square \\
& 871452.0 & --  &   &&   0.93168     && 82  && Triangle 
\\
\hline
\end{tabular}
\caption{\small
\label{tb:datafig22modified}
The same as Tables \ref{tb:datafig22} and \ref{tb:datafig22comment}
but the frequencies of the points for which 
$f_{\rm res} = 871434.4$ Hz in Table \ref{tb:datafig22comment}
have been shifted down by 0.4 Hz, to yield an `effective' frequency
(indicated by an asterisk in column 2).
Also rows for which $\sigma_p=6.0 \times 10^{-4}$ in Table \ref{tb:datafig22comment}
are indicated by a `$-$' in column 2.
The values in the above Table are plotted as the data in 
Figs.~\ref{fig:sweep-uncooled-2} and \ref{fig:sweep-uncooled_ramp}.
The last column is a key to indicate how the data is displayed in 
Figs.~\ref{fig:sweep-uncooled-2} and \ref{fig:sweep-uncooled_ramp}:
square -- points for which $f_{\rm res} = 871434.4$ Hz,
circle -- points for which $f_{\rm res} = 871434.0$ Hz
triangle -- points which were fitted with a simulation parameter value of 
$\sigma_p = 6.0 \times 10^{-4}$ (Fig.~\ref{fig:sweep-uncooled-2})
or $t_{\rm ramp}=0.02$ s (Fig.~\ref{fig:sweep-uncooled_ramp}).
}
\end{table}

\vfill\pagebreak

\vfill\pagebreak

\vfill\pagebreak

\vfill\pagebreak

\vfill\pagebreak

\vfill\pagebreak

\vfill\pagebreak

\vfill\pagebreak

\vfill\pagebreak

\vfill\pagebreak

\vfill\pagebreak

\vfill\pagebreak

\vfill\pagebreak

\vfill\pagebreak

\vfill\pagebreak

\vfill\pagebreak

\vfill\pagebreak

\vfill\pagebreak

\vfill\pagebreak

\vfill\pagebreak

\vfill\pagebreak

\vfill\pagebreak

\vfill\pagebreak

\vfill\pagebreak

\end{document}